\documentclass[a4paper,12pt]{article}

\usepackage{graphicx,array}
\usepackage{color}
\usepackage{latexsym}
\usepackage{amsthm}
\usepackage{amsmath}
\usepackage{amssymb}
\usepackage{empheq}
\usepackage{bm}

\usepackage{epsfig}
\usepackage{slashed}
\usepackage{psfrag}
\usepackage[svgnames]{xcolor}
\PassOptionsToPackage{caption=false}{subfig}
\usepackage{subcaption}
\usepackage{xfrac}
\usepackage{multirow}
\usepackage{booktabs}
\usepackage{cite}
\usepackage[normalem]{ulem}
\usepackage{jheppub} % for details on the use of the package, please see the JHEP-author-manual
\newcommand{\al}[1]{\begin{align}\begin{aligned} #1 \end{aligned}\end{align}}

\newcommand{\be}{\begin{equation}}
\newcommand{\ee}{\end{equation}}
\newcommand{\bea}{\begin{eqnarray}}
\newcommand{\eea}{\end{eqnarray}}
\newcommand{\bei}{\begin{itemize}}
\newcommand{\eei}{\end{itemize}}

\newcommand{\dd}{\text{d}}

\newcommand{\MPBH}{{M}_{\rm BH}}

\preprint{IPPP/24/25}

\title{Gravitationally produced Dark Matter and primordial black holes}

\author[a,b,c]{Enrico Bertuzzo,}
\author[d]{Yuber F. Perez-Gonzalez,}
\author[c]{Gabriel M. Salla,}
\author[c]{Renata Zukanovich Funchal}

\affiliation[a]{Dipartimento di Scienze Fisiche, Informatiche e Matematiche,
Universit\`a degli Studi di Modena e Reggio Emilia, Via Campi 213/A, I-41125 Modena, Italy}
\affiliation[b]{INFN sezione di Bologna, via
Irnerio 46, 40126 Bologna, Italy}
\affiliation[c]{Instituto de F\'isica, Universidade de S\~ao Paulo, C.P. 66.318, 05315-970 S\~ao Paulo, Brazil}
\affiliation[d]{Institute for Particle Physics Phenomenology, Durham University, South Road, Durham DH1 3LE, United Kingdom}

\emailAdd{enrico.bertuzzo@unimore.it}
\emailAdd{yuber.f.perez-gonzalez@durham.ac.uk}
\emailAdd{gabriel.massoni.salla@usp.br}
\emailAdd{zukanov@if.usp.br }

\abstract{
    We examine how the existence of a population of primordial black holes (PBHs) influences cosmological gravitational particle production (CGPP) for spin-0 and spin-1 particles.
    In addition to the known effects of particle production and entropy dilution resulting from PBH evaporation, we find that the generation of dark matter (DM) through CGPP is profoundly influenced by a possible era of PBH matter domination.
    This early matter dominated era results in an enhancement of the particle spectrum
    from CGPP. Specifically, it amplifies the peak comoving momentum $k_\star$ for spin-1 DM, while enhancing the plateau of the spectrum for minimally coupled spin-0 particles for low comoving momenta. 
    At the same time, the large entropy dilution may partially or completely compensate for the increase of the spectrum and strongly mitigates the DM abundance produced by CGPP. Our results show that, in the computation of the final abundance, CGPP and PBH evaporation cannot be disentangled, but the parameters of both sectors must be considered together to obtain the final result. 
    Furthermore, we explore the potential formation of PBHs from density fluctuations arising from CGPP and the associated challenges in such a scenario.
}

\begin{document} 
\maketitle

%%%%%%%%%%%%%%%%%%%%%%%%%%%%%%%%%%%%%%%%%%%%%%%%%%%%%%%%%%%%%%%%%%%%%%%%%%%%%%%%%%%%%%%%%%%%%%%%%%%%%%%%%%%%%%%%%%%%%%%%%%%%%%%%%%%%
\section{Introduction}
%%%%%%%%%%%%%%%%%%%%%%%%%%%%%%%%%%%%%%%%%%%%%%%%%%%%%%%%%%%%%%%%%%%%%%%%%%%%%%%%%%%%%%%%%%%%%%%%%%%%%%%%%%%%%%%%%%%%%%%%%%%%%%%%%%%%

It is a distinct possibility that Dark Matter (DM) only has gravitational interactions. This would explain why all the evidence we have for DM existence are either of cosmological or astrophysical nature, with any other type of search (direct and indirect detection, as well as collider) giving null results. If this is the case, the DM abundance could be set via Cosmological Gravitational Particle Production (CGPP), first proposed in Refs.\,\cite{Parker:1968mv,Parker:1969au,Parker:1971pt} and later explored, for instance, in Refs.\,\cite{Ford:1986sy,Chung:1998zb,Chung:2001cb,Chung:2004nh,Chung:2015tha,Graham:2015rva,Ema:2015dka,Ema:2016hlw,Ema:2019yrd,Chung:2018ayg,Bastero-Gil:2018uel,Ahmed:2020fhc,Kolb:2020fwh,Kolb:2021xfn,Ling:2021zlj,Arvanitaki:2021qlj,Kaneta:2022gug,Kolb:2022eyn,Hashiba:2022bzi,Redi:2022zkt,Gorghetto:2022sue,Kolb:2023dzp,Garcia:2023awt,Zhang:2023hjk,Zhang:2023xcd,Garcia:2023qab}, see also the recent reviews\,\cite{Ford:2021syk,Kolb:2023ydq}. In a nutshell, this production mechanism is based on the fact that, during and after inflation, the cosmological expansion causes the vacuum state to evolve with time, due to the time dependence of the DM field Hamiltonian. This implies that, even if we start in a ``zero particle'' vacuum very early during inflation, as time evolves the field will find itself in a state that is no longer the vacuum: a certain amount of particles will have been created. The phenomena continues until the particle becomes non-relativistic, after which the vacuum varies only adiabatically with time and CGPP ceases to be effective. It is important to observe that this production mechanism only requires the  expansion of the universe to work and is unavoidable, in the sense that it cannot be switched off.

Given that an abundance of DM particles will unavoidably be produced from CGPP, it is thus interesting to understand what happens to this mechanism once other gravitational phenomena that can occur in the early universe are considered. Since CGPP starts already during inflation, in this paper we will focus on what happens when it takes place in the presence of a population of Primordial Black Holes (PBHs) produced during inflation\,\cite{Hawking:1975vcx, Carr:1974nx,Carr:2009jm, Carr:2020gox,Heydari:2021gea,Heydari:2021qsr, Flores:2024eyy}. DM physics in the presence of PBHs has been considered many times in the past~\cite{Lennon:2017tqq,Hooper:2019gtx,Gondolo:2020uqv,Bernal:2020ili,Bernal:2020kse,Baldes:2020nuv,Masina:2021zpu,Sandick:2021gew,Bernal:2021bbv,Bernal:2021yyb,Cheek:2021odj,Cheek:2021cfe,Barman:2021ost,Bernal:2022oha,Barman:2024iht}. 
For what concerns DM, the most important effect is due to Hawking evaporation\,\cite{Hawking:1975vcx} and can be summarized as follows: ($i$) a population of DM particles is produced via evaporation and ($ii$) the DM abundance is diluted by the entropy injection due to evaporation. In the case of gravitationally produced DM, as we will see, a third phenomenon can take place: if the PBHs population dominates the expansion of the universe {\it before} DM becomes non-relativistic, then the additional phase of matter domination will qualitatively and quantitatively affect the final DM spectrum and abundance. This means that there can be a profound interplay between the physics involved in CGPP and PBHs. For simplicity, in what follows we will focus on bosonic DM, more precisely spin-0 and spin-1 DM, although our computations could be readily extended to fermionic DM.

The paper is organized as follows. In Sec.\,\ref{sec:GPP} we will discuss CGPP of bosonic DM, summarizing the main physics and equations. In Sec.\,\ref{sec:PBHs} we will then outline the main facts related to PBHs and their influence on DM physics. We will then present our main results in Sec.\,\ref{sec:results}, discussing in detail the interplay between CGPP and PBHs.
%, the final DM abundance and the main bounds that can be imposed on the parameter space. 
In particular, we will show how CGPP is affected, both at the level of number density spectrum and of DM abundance, when a phase of PBH domination takes place during the cosmological evolution. We then present the final DM abundance, taking into account the production via PBH evaporation and the corresponding bounds that can be imposed on the parameter space. Afterwards we discuss the possibility of PBHs being produced from density fluctuations coming from CGPP itself and the difficulties involved. Our conclusions will be given in Sec.\,\ref{sec:conclusions}. We also add two Appendices. In App.\,\ref{app:approximate_solutions_scalar}, we present approximate analytic results for CGPP of spin-0 DM, while in App.\,\ref{app:power_spectrum} we detail the computations of the power spectrum to determine the initial fraction of PBHs.
Throughout this manuscript, we consider natural units where $\hbar = c = k_{\rm B} = 1$, and define the Planck mass to be $m_{\rm PL}=1/\sqrt{8 \pi G}$, with $G$ being the gravitational constant.

%%%%%%%%%%%%%%%%%%%%%%%%%%%%%%%%%%%%%%%%%%%%%%%%%%%%%%%%%%%%%%%%%%%%%%%%%%%%%%%%%%%%%%%%%%%%%%%%%%%%%%%%%%%%%%%%%%%%%%%%%%%%%%%%%%%%
\section{Gravitational particle production of bosonic Dark Matter}\label{sec:GPP}
%%%%%%%%%%%%%%%%%%%%%%%%%%%%%%%%%%%%%%%%%%%%%%%%%%%%%%%%%%%%%%%%%%%%%%%%%%%%%%%%%%%%%%%%%%%%%%%%%%%%%%%%%%%%%%%%%%%%%%%%%%%%%%%%%%%%
Once Quantum Field Theory (QFT) is considered on a classical curved spacetime (as is the case in a 
Friedmann--Lemaitre--Robertson--Walker (FLRW) universe), the notion of ``vacuum'' of the theory is intrinsically ambiguous. 
The situation is analogous to the one of a quantum harmonic oscillator with time-dependent frequency\,\cite{Mukhanov:2007zz}.
The temporal evolution of the frequency implies that the expression of the coordinate operator in terms of creation and annihilation operators varies over time. Consequently, the vacuum state of the theory undergoes changes, as the operator responsible for annihilating the instantaneous vacuum state is not constant. 
The same reasoning applies to quantum fields, where the time evolution of the frequency term in the equation of motion of the mode functions is due to cosmological background evolution. We start by discussing the background cosmological evolution, and we then turn to a detailed treatment of scalar and vector CGPP.

%%%%%%%%%%%%%%%%%%%%%%%%%%%%%%%%%%%%%%%%%%%%%%%%%%%%%%%%%%%%%%%%%%%%%%%%%%%%%%%%%%%%%%%%%%%%%%%%%
\subsection{Background cosmological evolution}\label{sebsec:cosmo_evolution}
%%%%%%%%%%%%%%%%%%%%%%%%%%%%%%%%%%%%%%%%%%%%%%%%%%%%%%%%%%%%%%%%%%%%%%%%%%%%%%%%%%%%%%%%%%%%%%%%%

To be concrete, we will focus on quadratic (chaotic) inflation driven by a single scalar field $\Phi(x)$, the inflaton with mass $M$, with potential
\be\label{eq:chaotic_potential}
V(\Phi) = \frac{M^2}{2} \Phi^2 .
\ee
Although this model is experimentally excluded\,\cite{Planck:2018jri}, taking Eq.\,\eqref{eq:chaotic_potential} should be a reasonable approximation since we are mostly interested in the final stage of the inflaton evolution, i.e., when the field is close to the minimum of its potential and quite far away from the region experimentally probed by Cosmic Microwave Background (CMB) observables. 
Certainly, since we want to include the physical effects of a PBH population produced during inflation, some modification of Eq.\,\eqref{eq:chaotic_potential} will be needed in order to generate sufficiently large fluctuations that produce a PBH population.
We will explore this matter further in Sec.\,\ref{sec:results}, but we anticipate that, at least for the potentials we will consider (see Eq.\,\eqref{eq:inflaton_potential_PBHs}), the final effect on CGPP is small and can typically be neglected. 
With our assumptions, the inflaton field will, as usual, follow the equation of motion
\be\label{eq:inflaton_EqM}
\ddot{\Phi} + (3 H + \Gamma) \dot{\Phi}+ \frac{\partial V}{\partial\Phi} = 0,
\ee
where $\Gamma$ is the inflaton decay width, $H$ is the Hubble parameter, an overdot indicates a derivative with respect to cosmic time. The previous equation is valid during the quasi-de Sitter evolution, in the regime $\Gamma \ll H$, when the slow-roll parameters are small:
\be\label{eq:slow-roll-pars}
\epsilon = - \frac{\dot{H}}{H} \ll 1, ~~~~ \kappa = \left|\frac{\dot{\epsilon}}{H\,\epsilon}\right| \ll 1.
\ee
We will take the end of inflation to correspond to the moment in which $\epsilon = 1$. The values of the Hubble parameter and scale factor in this moment will be denoted by $H_e$ and $a_e$.\footnote{In order for $\epsilon=1$ to coincide with the moment at which $H=H_e$, the mass of the inflaton is fixed to be $M=2H_e$.} 
Once inflation ends and the inflaton begins oscillating around the minimum of its potential, we are in the reheating phase of the universe, described by the coupled set of equations\,\cite{Kolb:1990vq}
\be\label{eq:reheating_EoM}
\dot{\rho}_\Phi + 3 H \rho_\Phi = - \Gamma\, \rho_\Phi, ~~~~ \dot{\rho}_\text{R} + 4 H \rho_\text{R} = \Gamma\,\rho_\Phi,
\ee
where $\rho_\Phi$ and $\rho_\text{R}$ are the inflaton and radiation energy densities, respectively. The first of these equations is solved by
\be
\rho_\Phi(t) = \rho_\Phi(t_e) \left(\frac{a_e}{a}\right)^3 e^{-\Gamma(t-t_e)},
\ee
where $t_e$ is the time at which inflation ends. As for the second equation, it can be solved analytically in the regime in which the inflaton is still dominating the energy density budget (i.e. before radiation domination begins), giving
\be
\rho_\text{R}(t) = \frac{4 m_\text{PL}^2 \,\Gamma}{5t}\left[1-\left(\frac{t_e}{t}\right)^{5/3}\right].
\ee
If we now define that the period of radiation domination starts when 
% $\rho_\Phi/\rho_\text{tot} \lesssim 10^{-3}$
$\rho_\Phi/(\rho_\Phi+\rho_\text{R})\ll 1$
at a time $t_\text{RH}$, we obtain from the equations above that this happens provided that\,\footnote{To obtain the specific value $\Gamma t_\text{RH} \simeq 6$, we have taken $\rho_\Phi/(\rho_\Phi+\rho_\text{R}) \lesssim 10^{-3}$. This number is arbitrary and just a proxy for a small number. Different choices affect only mildly Eq.\,\eqref{eq:t}.}
\be\label{eq:t}
\Gamma \, t_\text{RH} \simeq 6.
\ee
We take the reheating temperature $T_\text{RH}$ as the temperature of radiation at the time defined by Eq.\,\eqref{eq:t}, from which we obtain a relation between $T_\text{RH}$ and $\Gamma$:
\be\label{eq:TRH}
\Gamma = 6 \sqrt{\frac{4\pi^2\, g_\star(T_\text{RH})}{90}} \frac{T_\text{RH}^2}{m_\text{PL}},
\ee
where $g_\star(T)$ is the effective number of relativistic degrees of freedom at temperature $T$ and $T_\text{RH}$ the temperature at $t_\text{RH}$. Apart from an $\mathcal{O}(1)$ factor, this agrees with the more standard definition of reheating temperature (see e.g.\,\cite{Kolb:1990vq}). In what follows, we will always fix the reheating temperature and use Eq.\,\eqref{eq:TRH} to infer the value of the inflaton decay width $\Gamma$ to be inserted in the equations of motion. Once the universe starts being dominated by radiation, we recover the radiation domination phase of the early universe in which Big-Bang Nucleosynthesis (BBN) and recombination happen. For the numerical analysis to be presented in Sec.\,\ref{sec:results}, we will solve numerically Eqs.\,\eqref{eq:inflaton_EqM} and\,\eqref{eq:reheating_EoM} and use them to compute the Hubble parameter $H(t)$ and the scale factor $a(t)$. Once these quantities are known, they can be inserted in the computation of the total number of particles produced via CGPP (see Eqs.\,\eqref{eq:EoM_scalar},\,\eqref{eq:frequency_T} and\,\eqref{eq:frequency_L} below).

%%%%%%%%%%%%%%%%%%%%%%%%%%%%%%%%%%%%%%%%%%%%%%%%
\begin{figure}[tb]
\centering
    \includegraphics[width=.48\textwidth]{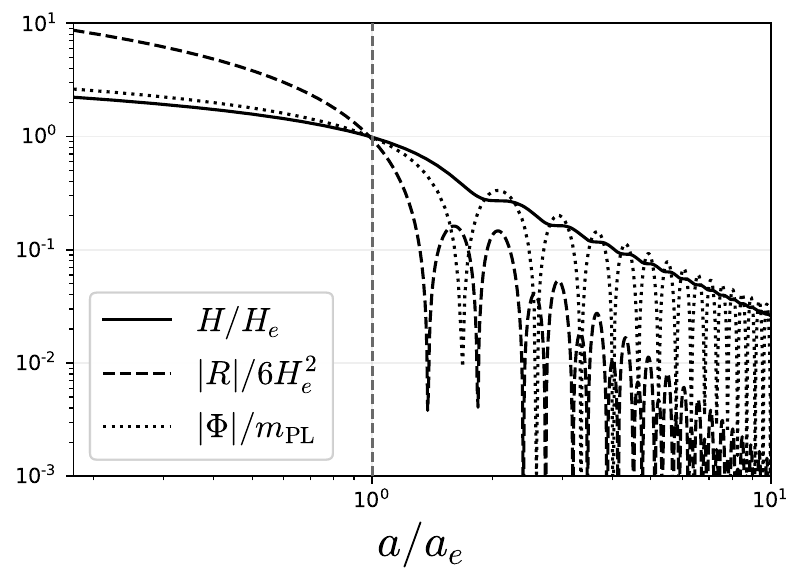}  
    \includegraphics[width=.48\textwidth]{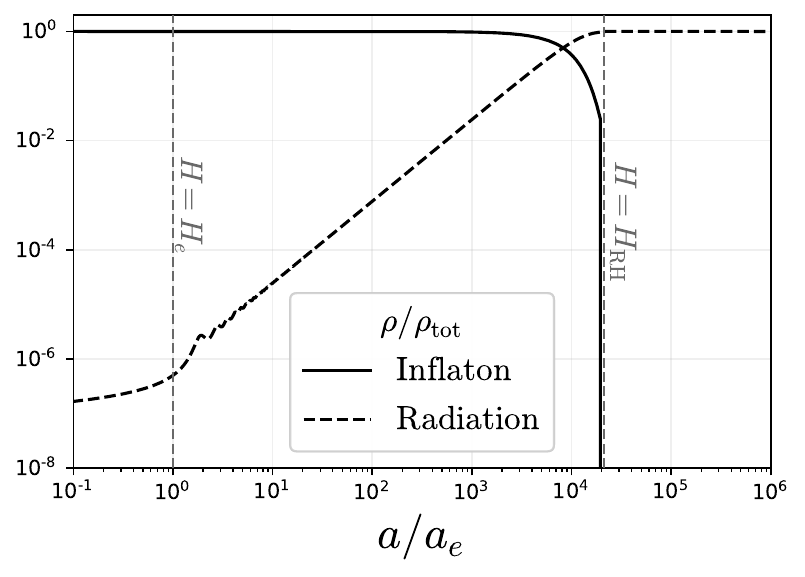}
    \caption{Background cosmological evolution for the quadratic inflationary potential\,\eqref{eq:chaotic_potential}. \textbf{Left panel:} Hubble parameter (solid, normalized to its value at the end of inflation $H_e$), the modulus of the Ricci scalar (dashed, normalized to $6 H_e^2$) and the absolute value of the inflaton field (dotted, normalized to $m_\text{PL}$) as a function of the scale factor normalized with respect to its value at the end of inflation. \textbf{Right panel:} inflaton and radiation energy density fractions as a function of the scale factor normalized by $a_e$. The two vertical dashed lines show the moments at which inflation and reheating ends. For both panels we have used $H_e=10^{13}$ GeV and $T_\text{RH}=10^{12}$ GeV.}
    \label{fig:StandardCosmo}
\end{figure}
%%%%%%%%%%%%%%%%%%%%%%%%%%%%%%%%%%%%%%%%%%%%%%%%
In Fig.\,\ref{fig:StandardCosmo} we show numerical results for the evolution of the universe as a function of the scale factor. On the left panel, we show the Hubble parameter, the Ricci scalar and the inflaton field around the end of inflation, where we notice the oscillating behaviour right after $a_e$, when reheating begins. On the right panel we show instead the energy densities of Standard Model (SM) radiation, which includes all the SM particles, and of the inflaton field, both normalized to the total energy density. We indicate the end of inflation by the leftmost gray vertical line, such that $H=H_e$, and the end of reheating by the rightmost line, where $H_\text{RH}=H(T_\text{RH})$.

\subsection{Cosmological gravitational particle production}
As already mentioned, CGPP is a consequence of the non-trivial evolution of the vacuum once QFT is considered on a classical curved spacetime. In the case under consideration, the vacuum evolution is due to the expansion of the FLRW universe, for which we will use the metric $g_{\mu\nu} = a^2(\eta) \text{diag}(1,-1,-1,-1)$, with $a(\eta)$ the scale factor as a function of conformal time, defined by $d\eta = dt/a$. 
Consider a quantum field corresponding to bosonic particles of a certain spin (in what follows, we will focus on the case of spin-0 and spin-1). The connection with the harmonic oscillator with a time-dependent frequency mentioned at the beginning of this section is most easily seen decomposing the canonically normalized quantum field as
\be\label{eq:decomposition}
\varphi^A(x) = \int\frac{d^3k}{(2\pi)^3} e^{-i\, \bm{k} \cdot \bm{x}} \left(v_k(\eta)\, \varepsilon^A_{\bm{k}}\, a_{\bm{k}} + v_{k}^*(\eta)\,\varepsilon^{A\star}_{-\bm{k}} \,a_{-\bm{k}}^\dag \right),
\ee
where $A$ is an appropriate Lorentz index and $\varepsilon_k^A$ are the corresponding polarization vectors (which, in the case of a spin-0 field, are trivial, while they are non-trivial for a spin-1 field). The mode functions $v_k(\eta)$ satisfy an equation of motion of the type
\be\label{eq:mode_equation}
v_k''(\eta) + \omega^2_k(\eta)\, v_k(\eta) = 0,
\ee
where a prime denotes derivative with respect to conformal time. This is precisely the equation of an harmonic oscillator with time-dependent frequency. The exact form of $\omega^2_k(\eta)$ depends on the spin of the field considered and will be shown later for the two cases under consideration. Given this time dependence, the decomposition in positive and negative energy solutions at a given time is not the same as at later times. This mismatch results in production of $\varphi$ quanta, i.e. we have CGPP. More concretely, this can be seen in the following way: take two times $\eta_1$ and $\eta_2 > \eta_1$ and denote the mode functions that solve Eq.\,\eqref{eq:mode_equation} and that minimize the energy at the two times as $v_{1k}$ and $v_{2k}$. Since these sets of mode functions are both basis, it is possible to write the expansion
\be
v_{1k}(\eta) = \alpha_k(\eta) v_{2k}(\eta) + \beta_k(\eta) v_{2k}^*(\eta),
\ee
which shows how the positive energy modes at $\eta_1$ are a combination of negative and positive energy modes at $\eta_2$. 
In general, if $\omega_k^2(\eta)$ is a rapidly varying function of time, the distinction between positive/energy modes varies drastically from one instant to the next and the notion of particle is not well-defined. Luckily, this is not the case for the FLRW universe described in Sec.\,\ref{sebsec:cosmo_evolution}, since although $\omega_k(\eta)$ is never constant, it changes slowly both in the far past and in the far future, in the sense that $\omega_k'/\omega_k^2 \ll 1$\,\cite{Kolb:2023ydq}.
This defines an approximate notion of vacuum (called adiabatic vacuum) based on the mode function
\be\label{eq:adiabatic_solution}
v_k^\text{ad}(\eta) = \frac{e^{-i \int^\eta d\eta' \omega_k(\eta')}}{\sqrt{2 \omega_k(\eta)}},
\ee
which corresponds to the lowest order approximation of a WKB solution of Eq.\,\eqref{eq:mode_equation}. For the FLRW universe under consideration, this can be easily seen to be true at early times, since, as we are going to discuss in detail later, the frequency is of the form $\omega_k^2(\eta) = k^2 + f(a, R)$ (with $f$ a function of the scale factor $a(\eta)$ and Ricci scalar $R(\eta)$ of appropriate dimensions) and, at early times, we have $f(a, R) \to 0$ for every mode $k$. This means that, at such early times, the notion of vacuum is defined by
\be\label{eq:bunch-davies_initial_cond}
v_k^\text{BD} = \frac{e^{- i k \eta}}{\sqrt{2k}},
\ee
and we will require the solutions of Eq.\,\eqref{eq:mode_equation} to satisfy these initial conditions (called Bunch--Davies initial conditions\,\cite{Bunch:1978yq}), that correspond to a state with no initial particles. At late times, we need the explicit form of $\omega_k^2$ to see that the frequency is indeed varying adiabatically and that the solutions of the mode function approach the adiabatic form of Eq.\,\eqref{eq:adiabatic_solution}. We will discuss this point later, when considering the specific cases of spin-0 and spin-1 fields. From our discussion, it follows immediately that CGPP starts to be effective when $f(a,R)\sim k^2$, while essentially no particles are created as long as $k^2 \gg f(a,R)$. When the adiabaticity condition starts to fail depends on the mode $k$ and on the explicit form of the frequency, in such a way that there will be modes for which CGPP starts towards the end of inflation and others for which it happens later. An analytic discussion of this point is presented in App.\,\ref{app:approximate_solutions_scalar} for the case of a conformally coupled scalar. Putting all together, we can resume the history of CGPP in the following way: under our assumptions, each mode starts, early during inflation, in the Bunch-Davies vacuum of Eq.\,\eqref{eq:bunch-davies_initial_cond}, when the vacuum is adiabatic and no relevant CGPP is taking place. As time passes, the adiabaticity condition fails, the frequency starts to evolve quickly and particle production starts. Depending on the mode, this may happen late during inflation or even after inflation has ended. CGPP continues to be effective as long as the vacuum is not varying adiabatically. Finally, at sufficiently late times during radiation domination, the frequency goes back to vary adiabatically, the vacuum is identified by the mode functions of Eq.\,\eqref{eq:adiabatic_solution} and CGPP ceases to be effective.

Once the solution of Eq.\,\eqref{eq:mode_equation} is known and the late-time adiabatic vacuum is used, the number of particles in a comoving volume can be computed according to\,\cite{Kolb:2023ydq}
\be\label{eq:number_density}
na^3 = \int \frac{d^3k}{(2\pi)^3} \left|\beta_k\right|^2,
\ee
where $n$ is the DM number density and $\beta_k$ are the so-called Bogoliubov coefficients, computed as\,\cite{Kolb:2023ydq}
\be\label{eq:Bogoliubov}
\left|\beta_k\right|^2 = \frac{\omega_k}{2} \left|v_k\right|^2 + \frac{\left| v_k'\right|^2}{2 \omega_k} - \frac{1}{2}.
\ee
Once more, we stress that the mode functions $v_k$ are computed imposing Bunch--Davies initial conditions (see Eq.\,\eqref{eq:bunch-davies_initial_cond}) for all the modes $k$, and they approach the adiabatic solution of Eq.\,\eqref{eq:adiabatic_solution} at late times. When this regime is reached, $|\beta_k|^2\simeq$ const and the number density of particles evolves as $a^{-3}$. 
We can then compute the number-to-entropy ratio $Y=na^3/sa^3$ at a moment after $na^3 = $ const and during radiation domination, where it makes sense to compute the entropy. If the comoving entropy $sa^3$ is conserved, we have that $d(sa^3)/d t =0$ and as a consequence $Y$ remains constant. Then, the abundance of DM today is
\be\label{eq:OmegaDM}
\Omega = \frac{m}{\rho_c} Y_0 s_0,
\ee
where $Y_0$ and 
$s_0\simeq 2\times 10^{-11}$ eV$^3$
are the number-to-entropy ratio and the entropy per comoving volume today, respectively, and the critical density is given by $\rho_c \simeq 8\times 10^{-11}h^2$ eV$^4$. To make explicit the dependence of the abundance on $H_e$ and $T_\text{RH}$, it is useful to rewrite it approximately as\,\cite{Kolb:2023ydq}
\be\label{eq:OmegaDM_approx}
\frac{\Omega h^2}{0.12}\simeq \frac{m}{H_e}\left(\frac{H_e}{10^{12}~\text{GeV}}\right)^2\left(\frac{T_\text{RH}}{10^{9}~\text{GeV}}\right)\frac{(na^3)_{H=m}/(a_eH_e)^3}{10^{-5}}\, .
\ee
If instead the comoving entropy is not conserved, as it can happen in the presence of PBHs, we will need to track its evolution in order to correctly compute $Y_0$ (see Eq.\,\eqref{eq:entropy_dilution}). 
We will show in Sec.\,\ref{sec:results} the numerical solution of Eq.\,\eqref{eq:mode_equation} and the final DM abundance for the two DM candidates considered.

\subsubsection{Spin-0 dark matter}
A scalar field $\phi(x)$ in a FLRW background is described by the action
\be
S_\text{scalar} = \int d^4x \sqrt{-g} \left(\frac{1}{2} g^{\mu\nu} \,\partial_\mu \phi \,\partial_\nu \phi - \frac{m^2}{2} \phi^2 +\frac{\xi}{2} R \phi^2 \right),
\ee
where $R$ is the Ricci scalar, $\xi$ a real parameter and $m^2$ is the mass parameter (assumed positive). Applying the field decomposition of Eq.\,\eqref{eq:decomposition} (with $\varepsilon = 1$) to the canonically normalized field $\varphi(x) = a(\eta) \phi(x)$, we obtain the mode equation shown in Eq.\,\eqref{eq:mode_equation} with frequency
\be\label{eq:EoM_scalar}
\omega_k^2(\eta) = k^2 + a^2 m^2 + \left(\frac{1}{6}-\xi\right)a^2R.
\ee
The form of the frequency is such that, for every mode $k$, there is a time sufficiently early in the history of the universe for which $\omega_k^2 \to k^2$, which is the condition of adiabaticity of the vacuum at early times discussed above, that allows us to use Bunch--Davies initial conditions for the mode functions. At late times, the term $a^2 R$ decreases as $a^{-1}$ (during matter domination) or vanishes (during radiation domination), in a such a way that, for each mode $k$, there will be a time sufficiently late in the history of the universe for which the $a^2 m^2$ term will dominate over the $k^2$ term. In this regime, the adiabaticity condition $\omega_k'/\omega_k^2 \ll 1$ becomes
\be\label{eq:adiabatic_condition_scalar}
\frac{\omega_k'}{\omega_k^2} \simeq \frac{a'}{a^2\, m} = \frac{H(\eta)}{m} \ll 1.
\ee
We thus see that we can start talking about a (very) rough notion of adiabaticity when $H = m$ (which, in physical terms, corresponds to the moment in which the comoving Compton wavelenght $1/am$ of the scalar particle is equal to the comoving Hubble horizon $1/aH$) and that this notion becomes more and more accurate as time passes. As explained above, in this regime $|\beta_k|^2 \simeq$ const and we can compute the final DM abundance according to Eq.\,\eqref{eq:OmegaDM}.\,\footnote{Numerically, it is not enough to stop the evolution of the modes at $H=m$, because $|\beta_k|^2$ is still oscillating. In order to the Bogoliubov coefficients to converge and reach a stable value, we stop the evolution when $H\sim m/100$.} As we are going to see in Sec.\,\ref{sec:results}, the overall DM physics depends crucially on when the adiabaticity condition happens, i.e. if $H=m$ happens during a phase of matter or radiation domination. 

Equation\,\eqref{eq:mode_equation} with frequency\,\eqref{eq:EoM_scalar} must be solved numerically, since no general analytic solution is known for all times. This is the strategy we follow in Sec.\,\ref{sec:results}.
In spite of this, it is possible to find approximate analytical solutions for different values of the parameters $k$ and $m$. We present our analytical results in App.\,\ref{app:approximate_solutions_scalar}, focusing on the case $\xi=1/6$ for simplicity.

An important point that must be discussed for gravitationally produced spin-0 DM are the limits from isocurvature perturbations. In general, the spin-0 field will inherit the usual adiabatic inflaton fluctuations, but may have additional large isocurvature fluctuations on small scales that are excluded by CMB measurements. This can be intuitively understood observing that the power spectrum of the scalar field will be proportional to the spectrum $n_k \equiv k^3 |\beta_k|^2/(2\pi^2)$, in such a way that, if such spectrum is large for small $k$, the contributions to isocurvature perturbations will, in general, be too large to be compatible with CMB data\,\cite{Chung:2004nh,Chung:2015tha,Garcia:2023qab}. Approximate analytic solutions of the mode equation\,\eqref{eq:mode_equation} show that, for small $k$, we approximately have $n_k \sim$ const for $\xi = 0$ and $n_k \sim k^2$ for $\xi=1/6$\,\cite{Garcia:2023qab,Kolb:2023ydq}, in such a way that it is expected that a minimally coupled scalar with $\xi = 0$ is excluded by isocurvature perturbations, while a conformally coupled scalar with $\xi=1/6$ is not. A detailed numerical analysis\,\cite{Garcia:2023qab} shows that isocurvature perturbations limits are not effective as soon as $\xi \gtrsim 1/48$. Instead of showing our results for different values of $\xi$, in what follows we will simply show the two representative cases $\xi = 0$ (minimal coupling) and $\xi = 1/6$ (conformal coupling), with the understanding that the interesting physics will roughly happen between these two values.

\subsubsection{Spin-1 dark matter}
%%%%%%%%%%%%%%%%%%%%%%%%%%%%%%%%%%%%%%%%%%%%%%%%%%%%%%%%%%%%%%%%%%%%%%%%%%%%%%%%%%%%%%%%%%%%%%%%%
Turning to spin-1 DM, the action is given by
\be
S_\text{vector} = \int d^4x \sqrt{-g} \left(-\frac{1}{4} g^{\mu\alpha} g^{\nu\beta} F_{\mu\nu} F_{\alpha\beta} + \frac{m^2}{2} g^{\mu\nu} A_\mu A_\nu \right),
\ee
where we neglect possible couplings between the DM and the Ricci tensor and scalar. The situation is now more involved with respect to the case of the scalar field, since the 4-vector $A_\mu$ contains, in addition to the physical degrees of freedom (two transverse fields $\bm{A}^T_i$ ($i=1,2$) and one longitudinal field $A_L$), also the non-dynamical temporal component. This can be eliminated from the action using its equation of motion\,\cite{Graham:2015rva,Ema:2019yrd,Ahmed:2020fhc,Kolb:2020fwh}. The decomposition in terms of creation and annihilation operators can be done directly for the transverse fields $\bm{A}^T_i$, while for the longitudinal field it is convenient to introduce the auxiliary field
\be
\chi_L = \frac{a\, m}{\sqrt{k^2 + a^2\, m^2}}\, A_L.
\ee
The decomposition of $\bm{A}^T_i$ and $\chi_L$ is now analogous to the one in Eq.\,\eqref{eq:decomposition}. Indicating with $v_k^{T,i}(\eta)$ and $v_k^L(\eta)$ the mode functions for $\bm{A}^T_i$ and $\chi_L$, respectively, we obtain the equations of motion
\be\label{eq:EoM_vector}
(v_k^{T,i})'' + \omega_{T,k}^2 \,v_k^{T,i} = 0, ~~~~~~ (v_k^{L})'' + \omega_{L,k}^2 \,v_k^{L} = 0,
\ee
with frequencies given by
\be\label{eq:frequency_T}
\omega_{T,k}^2 = k^2 + a^2\, m^2,
\ee
and
\be\label{eq:frequency_L}
\omega_{L,k}^2 = k^2 + a^2\, m^2 +\frac{a^2k^2}{k^2 + a^2\,m^2} \left[\frac{R}{6} + \frac{3\,a^2 m^2H^2}{k^2+a^2\,m^2}  \right].
\ee
We see that while the frequency of the transverse modes is identical to the one of a conformally coupled scalar, the one of the longitudinal mode is considerably more complicated. This is due to the non-trivial metric appearing in the mass term $g^{\mu\nu} A_\mu A_\nu$ for spin-1 DM, as opposed to the simple $m^2 \varphi^2$ term of the scalar case. The complicated expression for $\omega_k^L$ has an important consequence for CGPP: since $\omega_{L,k}^2$ can be negative, for certain values of the parameters there may be an exponential growth of the mode functions and, as a consequence, of the number of gravitationally produced DM particles. Most of the DM will thus reside in the longitudinal mode of the vector, with only a subdominant component made up of transverse modes. 

As intuitively clear, for the transverse mode the notion of adiabatic vacuum is identical to the one already discussed for the scalar case (see Eq.\,\eqref{eq:adiabatic_condition_scalar}). For the longitudinal mode, things are apparently much more complicated. The complication is only apparent once we realize that $\omega_{L,k}^2$ depends on three combinations of parameters: $k^2$, $a^2 H^2$ and $a^2 m^2$\,\cite{Ahmed:2020fhc}. For each mode $k$, at sufficiently early times we still obtain $\omega_{L,k}^2 \simeq k^2$ and hence we can still use Bunch--Davies initial conditions for the mode functions. On the other hand, at sufficiently late times (when $a m \gg aH, k$), the frequency can be approximated as $\omega_{L,k}^2 \simeq a^2 m^2$, giving the same adiabaticity condition as in the scalar case, see Eq.\,\eqref{eq:adiabatic_condition_scalar}.

As in the scalar case, no analytic solutions of the equations of motions\,\eqref{eq:EoM_vector} are available for all times. Approximate analytic solutions have been investigated in Refs.\,\cite{Graham:2015rva, Ema:2019yrd,Ahmed:2020fhc, Kolb:2020fwh}. Once the mode functions are known, the number of gravitationally produced DM particles is still given by Eqs.\,\eqref{eq:number_density}-\eqref{eq:Bogoliubov}, while to compute the abundance in the absence of PBHs we can use Eq.\,\eqref{eq:OmegaDM} and\,\eqref{eq:OmegaDM_approx}. We will show in Sec.\,\ref{sec:results} numerical solutions of Eq.\,\eqref{eq:EoM_vector}, while we refer the reader to Refs.\,\cite{Graham:2015rva, Ema:2019yrd,Ahmed:2020fhc, Kolb:2020fwh} for the approximate expressions.

We conclude this section stressing that, unlike what happens in the scalar case, spin-1 DM never generates too large isocurvature perturbations because $n_k$ is never large at small $k$. This important result has been discussed in detail in Refs.\,\cite{Graham:2015rva,Ahmed:2020fhc}.

%%%%%%%%%%%%%%%%%%%%%%%%%%%%%%%%%%%%%%%%%%%%%%%%%%%%%%%%%%%%%%%%%%%%%%%%%%%%%%%%%%%%%%%%%%%%%%%%%%%%%%%%%%%%%%%%%%%%%%%%%%%%%%%%%%%%
\section{Primordial black holes}\label{sec:PBHs}
%%%%%%%%%%%%%%%%%%%%%%%%%%%%%%%%%%%%%%%%%%%%%%%%%%%%%%%%%%%%%%%%%%%%%%%%%%%%%%%%%%%%%%%%%%%%%%%%%%%%%%%%%%%%%%%%%%%%%%%%%%%%%%%%%%%%

Various mechanisms have been proposed for generating a population of black holes in the early Universe, see, e.g., Refs.\,\cite{Carr:2009jm, Carr:2020gox, Flores:2024eyy} for reviews of such mechanisms. 
Among these, the collapse of large density fluctuations is a prominent mechanism connecting PBH formation with the inflationary paradigm\,\cite{Hawking:1975vcx, Carr:1974nx}. 
Essentially, when large density fluctuations reenter the horizon after inflation, they may collapse into a black hole if they exceed a certain threshold\,\cite{Press:1973iz}. 
Consequently, an inflationary model must account for the origin of such large fluctuations while simultaneously reproducing the observed scale-invariant CMB power spectrum. 
To achieve this, it has been demonstrated that the inflaton potential must include an additional flat region, resulting in a peak in the power spectrum\,\cite{Ivanov:1994pa,Inomata:2017okj,Ballesteros:2017fsr}.
If such flatness exists in the inflaton potential either by construction or because of the existence of an inflection point, the density fluctuations will collapse when their comoving wavelength $k$ is of the same order of the Hubble scale.
The PBH mass will then be proportional to the particle horizon mass via\,\cite{Carr:2009jm, Carr:2020gox}
\begin{align}\label{eq:PBH_formation}
  M_{\rm BH} = \frac{4\pi}{3}\, \gamma\, \frac{\rho}{H^3}\,,
\end{align}
with $\rho$ the total energy density at horizon reentering, and $\gamma$ the gravitational collapse factor. Assuming the Universe to be radiation dominated at that point, one can connect the PBH mass with the $k$ value at formation via\,\cite{Inomata:2017okj,Ballesteros:2017fsr}
\begin{align}
    M_{\rm BH}(k) &= \frac{\gamma}{2G}\frac{1}{H_{\rm eq}} \left(\frac{g_\star(T_{\rm eq})}{g_\star(T_{\rm f})}\right)^{1/6}  \left(\frac{k_{\rm eq}}{k}\right)^{2},\notag\\
    &\sim  10^7~{\rm g} \left(\frac{\gamma}{0.2}\right) \left(\frac{106.75}{g_\star(T_{\rm f})}\right)^{1/6}  \left(\frac{2.21\times 10^{19}~{\rm Mpc^{-1}}}{k}\right)^{2},
\end{align}
where the subscripts ``eq'' indicate quantities computed at matter-radiation equality, and those with ``f'' correspond to quantities at PBH formation.
Moreover, we can relate the formation temperature to the initial PBH mass, as
\begin{align}~\label{eq:temp_BHMass}
    M_{\rm BH}(T_{\rm f}) &= 3 \gamma \left(\frac{160}{g_\star (T_{\rm f})}\right)^{1/2} \frac{m_{\rm PL}^3}{T_{\rm f}^2}\notag\\
    &\sim 1.89\times 10^7~{\rm g}\left(\frac{\gamma}{0.2}\right) \left(\frac{10^{12}~{\rm GeV}}{T_{\rm f}}\right)^2.
\end{align}
The collapse is described according to the Press-Schechter model\,\cite{Press:1973iz}. In this model, the initial PBH energy density fraction 
\begin{align}\label{eq:beta_in}
 \beta(M_{\rm BH}) \equiv \frac{\rho_{{\rm PBH}} \left(T_{\rm f}\right)}{\rho\left(T_{\rm f}\right)}\,,
\end{align}
is linked to the probability of an overdensity $\delta$ to collapse forming a PBH. 
$\beta(M_{\rm BH})$ will depend on the equation-of-state of the Universe when the overdensity reenters the horizon.
If the Universe is radiation dominated, the density perturbation needs to exceed a certain threshold $\delta_c$ to collapse. Assuming a Gaussian statistics for the density perturbations\,\cite{Young:2014ana,Inomata:2017okj,Ballesteros:2017fsr}, we have
\begin{align}
    \beta(M_{\rm BH}) = \frac{1}{\sqrt{2\pi \sigma^2(M_{\rm BH})}}\int_{\delta_c}^\infty\, d\delta \exp\left(-\frac{\delta^2}{2\sigma^2(M_{\rm BH})}\right)
\end{align}
where $\sigma^2(M_{\rm BH})$ is the standard deviation of the coarse-grained density contrast for a PBH with mass $M_{\rm BH}$\,\cite{Young:2014ana}.
The variance $\sigma^2(M_{\rm BH}(k))$ for a general equation-of-state parameter $\omega$ is given by\,\cite{Young:2014ana}
\begin{align}
    \sigma^2(M_{\rm BH}(k)) = \int\,\frac{dq}{q} {\cal P}_\delta(q) W\left(\frac{q}{k}\right)^2 = \left(\frac{2(1+\omega)}{5+3\omega}\right)^2\int\,\frac{dq}{q} \left(\frac{q}{k}\right)^4 {\cal P}_{\cal R}(q) W\left(\frac{q}{k}\right)^2,
\end{align}
being ${\cal P}_\delta, {\cal P}_{\cal R}$ the density and comoving curvature power spectra, respectively, and $W(x)$ is a Gaussian smoothing window function.

Although there are some caveats in the description above, such as the validity of the Gaussian statistics for the density perturbation, the exact value of the threshold value of $\delta_{c}$ or the exact form of the window function, we observe that the comoving curvature power spectra, which depends on the inflaton potential, is crucial to predict the PBH mass and abundance. 
Obtaining such power spectra requires a detailed numerical simulation that solves the Mukhanov--Sakaki equation\,\cite{Sasaki:1986hm,Mukhanov:1988jd}, as seen in Refs.\,\cite{Sasaki:2018dmp, Ballesteros:2017fsr, Mishra:2019pzq}. 
For sake of completeness, we provide details on the numerical approach performed to solve the Mukhanov--Sakaki equation and the computation of the curvature power spectrum in App.\,\ref{app:power_spectrum}.
However, as our main focus is on understanding how additional features of the inflaton potential could alter bosonic DM production via CGPP, we first need to consider the modifications to the potential required for PBH formation.
For this, we follow the procedure established in Ref.\,\cite{Mishra:2019pzq}, where a local bump/dip is added to the potential in the form
\begin{align}\label{eq:inflaton_potential_PBHs}
    V(\Phi) \to V(\Phi)(1+\varepsilon(\Phi)),
\end{align}
where $\varepsilon(\Phi)$ is parametrized to have a Gaussian form 
\begin{align}
    \varepsilon(\Phi) = A \exp\left[-\frac{(\Phi - \Phi_0)^2}{2\sigma_\Phi^2}\right],
\end{align}
where $\{A, \Phi_0, \sigma_\Phi\}$ characterize the height, position and width of the bump\,\cite{Mishra:2019pzq}. 

\begin{figure}[t!]
\centering
    \includegraphics[width=\linewidth]{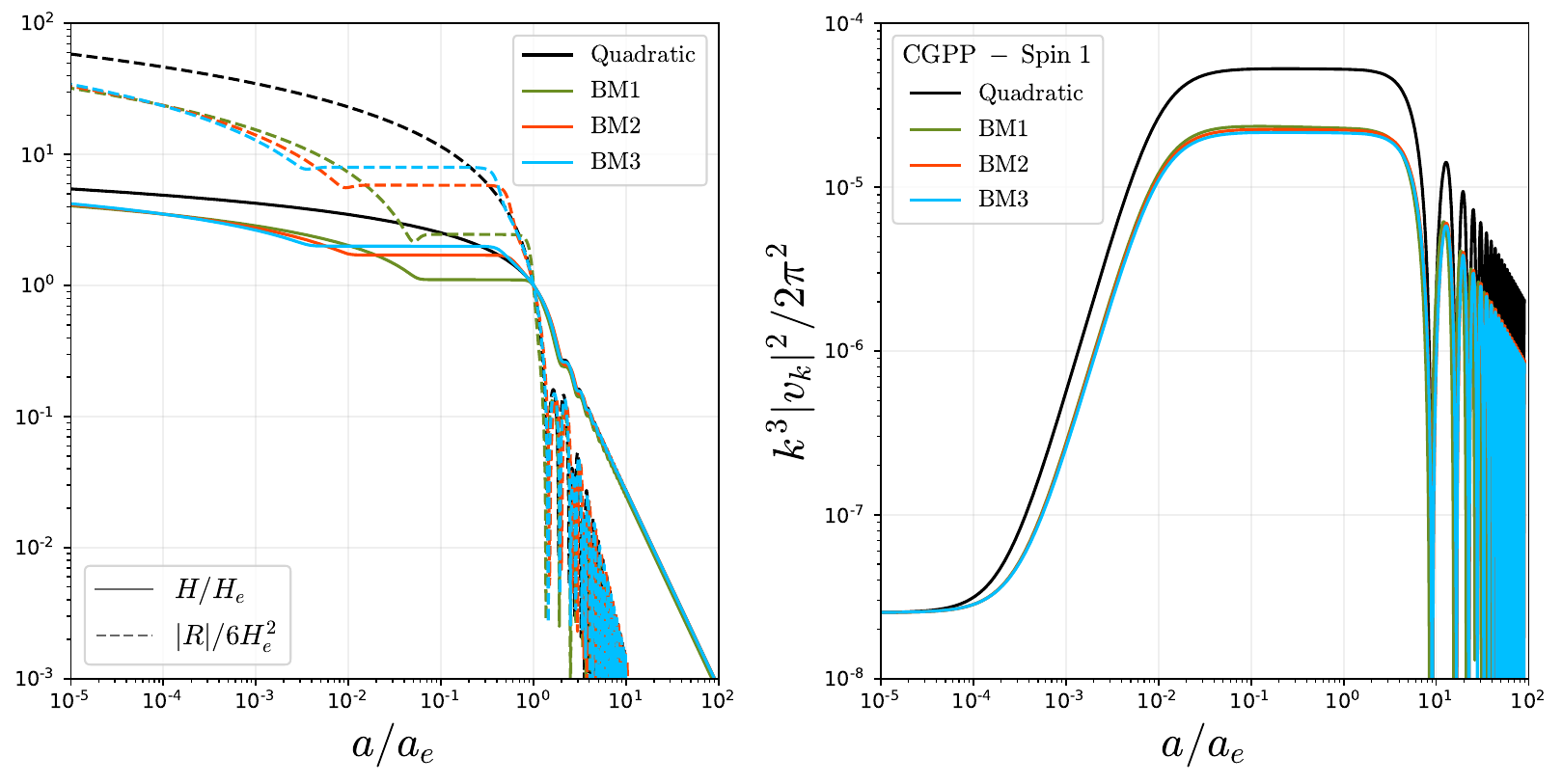}
    \caption{Predictions for different inflationary potentials, choosing $H_e=10^{13}$ GeV and $T_\text{RH}=10^{12}$ GeV. \textbf{Left panel:} Hubble parameter (solid) and Ricci scalar (dashed), both normalized as in Fig.\,\ref{fig:StandardCosmo}. In black we assume the quadratic potential of Eq.\,\eqref{eq:chaotic_potential} and in color we have curves for different benchmark models defined in Tab.\,\ref{tab:exam_pbh}. \textbf{Right panel:} Solution of Eq.\,\eqref{eq:EoM_vector} for a spin-1 particle, for the mode with $k = 10^{-3}\, a_eH_e$.}
    \label{fig:evol_USR}
\end{figure}

It is worth noting that the characteristics of the ``speed breaker'' dictate the peak properties in the curvature power spectrum, and consequently, the PBH mass and distribution. Therefore, to generate PBHs that constitute the DM, i.e., PBHs with masses $M_{\rm BH} \sim 10^{18}{\rm g}-10^{20}{\rm g}$, the peak must be situated far from the potential's minimum\,\cite{Mishra:2019pzq}. 
Consequently, we anticipate that dark particles produced via CGPP will remain unaffected by this additional feature (we remind the reader that, according to our discussion in Sec.\,\ref{sec:GPP}, early during inflation the evolution is adiabatic and no relevant CGPP takes place, so that the phenomenon is expected to happen for most modes either towards the end of inflation or after). 
However, if the feature is in close proximity to the minimum, we may expect alterations to the particle generation from the vacuum. 
In Tab.\,\ref{tab:exam_pbh} we show three benchmark choices of $\{A,\Phi_0,\sigma_\Phi\}$ that result in PBHs with masses $M_{\rm BH} \sim 10^{3}{\rm g}-10^{6}{\rm g}$.\footnote{Note that we quote the values of $A$ up the sixth decimal place such that the initial PBH abundance is $\beta \gtrsim 10^{-5}$. As noted in the literature\,\cite{Ballesteros:2017fsr,Mishra:2019pzq}, this high level of fine tuning is required to produce a significant PBH population.} We present in Fig.\,\ref{fig:evol_USR} (left panel) the Hubble parameter and Ricci scalar for the quadratic potential of Eq.\,\eqref{eq:chaotic_potential} and for the three benchmarks described in the table. In the right panel we instead present the effects of the different potential choices on the solution of the mode equation of Eq.\,\eqref{eq:EoM_vector}. We will show in the next section the final effect of such additional speed bump on the DM abundance.

%%%%%%%%%%%%%%%%%%%%%%%%%%%%%%%%%%%%%%%%%
\begin{table}[t]
\caption{Benchmark values of the parameters of the inflaton potential with a period of ultra-slow-roll that produce PBHs.
\label{tab:exam_pbh}}
    \centering
    \begin{tabular}{lcccc}
        \toprule\toprule
         & $A$ & $\Phi_0/m_{\rm PL}$ & $\sigma_\Phi/m_{\rm PL}$ &  $M_{\rm BH}~[{\rm g}]$  \\ \midrule\midrule
       BM1 & $0.582578$ & $1.25$ & $\sqrt{0.02}$ & $10^3$ \\ \midrule
       BM2 & $0.300557$ & $2.00$ & $\sqrt{0.02}$ & $10^5$ \\ \midrule
       BM3 & $0.256025$ & $2.25$ & $\sqrt{0.02}$ & $10^6$ \\ \midrule
          \bottomrule
    \end{tabular}
\end{table}
%%%%%%%%%%%%%%%%%%%%%%%%%%%%%%%%%%%%%%%%%

Once the population has formed, the PBHs begin emitting all degrees of freedom present in nature, depending on their initial characteristics\,\cite{Hawking:1975vcx}. 
Although PBHs could potentially have significant angular momentum, for simplicity, we assume here that they are Schwarzschild throughout their entire lifespan.
The emission rate of a particle $i$ with spin $s_i$ and internal degrees of freedom $g_i$ is determined by the Hawking spectrum
\begin{align}\label{eq:emission_rate}
 \frac{d^{2}N_{i}}{dt\, dE} = \frac{g_{i}}{2\pi}\frac{\Gamma_i(\MPBH,E)}{e^{E/T_{{\rm BH}}}-\left(-1\right)^{2s_{i}}}\,,
\end{align}
where the PBH temperature is related to its mass via
\begin{align}\label{eq:T_BH}
    T_{\rm BH}= \frac{1}{8\pi G M_{\rm BH}} \sim 10^3~{\rm TeV} \left(\frac{10^{7}~{\rm g}}{M_{\rm BH}}\right).
\end{align}
The Hawking spectrum differs from that of a blackbody because of the curved spacetime properties around the PBH. When a particle is emitted, it encounters an effective potential barrier, which could cause it to be reabsorbed by the black hole. Therefore, it is necessary to account for the probability of a particle reaching spatial infinity by incorporating the spin-dependent absorption probabilities $\Gamma_i(\MPBH,E)$ in Eq.\,\eqref{eq:emission_rate}. 
From simple energy conservation arguments, we can estimate the PBH mass loss rate\,\cite{MacGibbon:1990zk, MacGibbon:1991tj,Cheek:2021odj}
\begin{align} \label{eq:mass_loss}
 \frac{d\MPBH}{dt} = -\sum_{i} \int_{m_{i}}^{\infty} \frac{d^{2}N_{i}}{dt\, dE}\, E\, dE = - \varepsilon(\MPBH)\, \frac{m_{\rm PL}^4}{\MPBH^2}\,,
\end{align}
with $\varepsilon(\MPBH)$ the evaporation function that contains the information of the degrees of freedom that can be emitted for a given PBH mass, see, e.g., Refs.\,\cite{MacGibbon:1990zk,Cheek:2021odj}.

However, there remain several unresolved questions concerning evaporation phy\-sics, including the information paradox\,\cite{Hawking:1976ra,Almheiri:2020cfm,Buoninfante:2021ijy} and the thermal nature of the Hawking spectrum after the Page time\,\cite{Page:1993wv,Page:2013dx}. 
Resolving these issues could result in modifications to the PBH time evolution as described above, or even challenge the validity of the semi-classical approximation used to derive the Hawking spectrum, see however~\cite{Bardeen:1981zz,Massar:1994iy,Brout:1995rd}. 
Since the extent of such modifications is unclear, we adopt an agnostic approach and assume that the PBH time evolution follows the mass loss rate in Eq.\,\eqref{eq:mass_loss}, while also assuming the validity of the semi-classical approximation until near the Planck scale.

One significant consequence of the PBH population's presence in the early Universe is the potential for an early matter-dominated era following reheating. 
This arises from the behavior of PBHs, which act as pressure-less matter and redshift much more slowly than radiation, $\rho_{\rm PBH} \propto a^{-3}$. 
Consequently, even a small initial population of PBHs could eventually dominate the Universe's energy density.

In general, inflationary PBH are expected to have an extended mass distribution parametrized via $\beta (\MPBH)$, depending on the speed breaker feature.
In what follows, we will assume that all PBH have the same mass, i.e., the PBH population follows a monochromatic distribution for the sake of simplicity.
In this case, we can estimate the minimal initial density that would lead to a PBH-dominated early Universe~\cite{Hooper:2019gtx,Lunardini:2019zob}
\begin{align}\label{eq:beta_crit}
    \beta \gtrsim \beta_{c} \equiv 2.5\times 10^{-13} \left(\frac{g_\star(T_{\rm f})}{106.75}\right)^{-1/4}\left(\frac{M_{\rm BH}}{10^8~{\rm g}}\right)^{-1}.
\end{align}
The Friedmann-Boltzmann equations for an Universe containing a PBH population characterized with an energy density $\rho_{\rm PBH}$ after reheating are given by
\begin{subequations}\label{eq:FBEqs}
\begin{align}
    \dot{\rho}_{\rm R} + 4H \rho_{\rm R} &= -\left.\frac{d \ln \MPBH}{d t}\right|_{\rm SM}\rho_{\rm PBH}\,,\\
    \dot{\rho}_{\rm PBH} + 3H \rho_{\rm PBH} &= \frac{d \ln \MPBH}{d t}\rho_{\rm PBH}\,,
\end{align}
\end{subequations}
where subscript ``SM'' denotes that we are considering only the SM contribution to the evaporation to reheat the thermal bath. 
Because of this particle injection, entropy is no longer conserved. 
To determine the entropy dilution to any decoupled species' abundance, we track the entropy density through the equation\,\cite{Lunardini:2019zob}
\begin{align}\label{eq:entropy_dilution_eq}
    \dot{s}_{\rm R} + 3Hs_{\rm R} = -\frac{1}{T}\left.\frac{d \ln \MPBH}{d t}\right|_{\rm SM}\rho_{\rm PBH}.
\end{align}
We can estimate the entropy dilution factor considering energy conservation before and after PBH evaporation, see e.g.~Ref.\,\cite{Bernal:2022pue},
\begin{align}\label{eq:entropy_dilution}
     \frac{s(\tilde T)}{s(T_{\rm ev})} = \left(\frac{\tilde{T}}{T_{\rm ev}}\right)^{3} = \left(1+\frac{\beta T_{\rm f}}{T_{\rm ev}}\right)^{3/4},
\end{align}
where $T_{\rm ev} (\tilde T)$ is the plasma temperature right before (after) evaporation.
The effect of such a dilution will be crucial for the final dark boson abundance once we allow the presence of a PBH population during CGPP.
%%%%%%%%%%%%%%%%%%%%%%%%%%%%%%%%%%%%%%%%%%%%%%%%
%%%%%%%%%%%%%%%%%%%%%%%%%%%%%%%%%%%%%%%%%%%%%%%%

In the scenario where the initial fraction and PBH mass are treated as free parameters, allowing for various formation mechanisms beyond the inflationary case discussed earlier, constraints can be placed based on the potential implications of this PBH population throughout the history of the Universe\,\cite{Carr:2020gox, Carr:2020xqk}.
Such bounds will depend on whether the PBHs have evaporated.
Since CGPP occurs in the very early Universe, our focus will be on PBHs that evaporated prior to BBN, with $\MPBH\lesssim 10^9$\,g.
There exist some constraints on such PBH population, but they are highly model-dependent\,\cite{Carr:2020gox, Carr:2020xqk, Lunardini:2019zob}. 
However, recent constraints have been established by considering the gravitational waves (GWs) produced as a consequence of the presence of the PBH population, the potential early matter-dominated era, and the rapid transition to radiation domination after evaporation\,\cite{Papanikolaou:2020qtd,Domenech:2020ssp}.
Specifically, to avoid a backreaction issue, it is crucial that the energy contained in GWs does not surpass that of the background Universe\,\cite{Papanikolaou:2020qtd}. Additionally, modifications to BBN predictions due to the energy density stored in GWs can be circumvented if\,\cite{Domenech:2020ssp}
\begin{equation} \label{eq:GW}
  \beta \lesssim 1.1 \times 10^{-6} \left(\frac{\gamma}{0.2}\right)^{-1/2} \left(\frac{M_{\rm BH}}{10^4~\text{g}}\right)^{-17/24}.
\end{equation}
Further constraints are imposed on the DM particles emitted from the Hawking evaporation of the PBH population. 
Specifically, given that the PBH temperature can significantly exceed the DM mass, the emitted particles may be highly boosted, resulting in DM that is too warm during structure formation, thus violating Lyman-$\alpha$ constraints\,\cite{Bode:2000gq,Boyarsky:2008xj,Baldes:2020nuv,Baur:2017stq,Auffinger:2020afu, Masina:2021zpu, Cheek:2021odj, Cheek:2022mmy}. We will come back to this bound in Sec.\,\ref{sec:results}.

%%%%%%%%%%%%%%%%%%%%%%%%%%%%%%%%%%%%%%%%%%%%%%%%%%%%%%%%%%%%%%%%%%%%%%%%%%%%%%%%%%%%%%%%%%%%%%%%%%%%%%%%%%%%%%%%%%%%%%%%%%%%%%%%%%%%
\section{Results}\label{sec:results}
%%%%%%%%%%%%%%%%%%%%%%%%%%%%%%%%%%%%%%%%%%%%%%%%%%%%%%%%%%%%%%%%%%%%%%%%%%%%%%%%%%%%%%%%%%%%%%%%%%%%%%%%%%%%%%%%%%%%%%%%%%%%%%%%%%%%
%%%%%%%%%%%%%%%%%%%%%%%%%%%%%%%%%%%%%%%%%%%%%%%%
\begin{figure}[t!]
\centering
    \includegraphics[width=.9\linewidth]{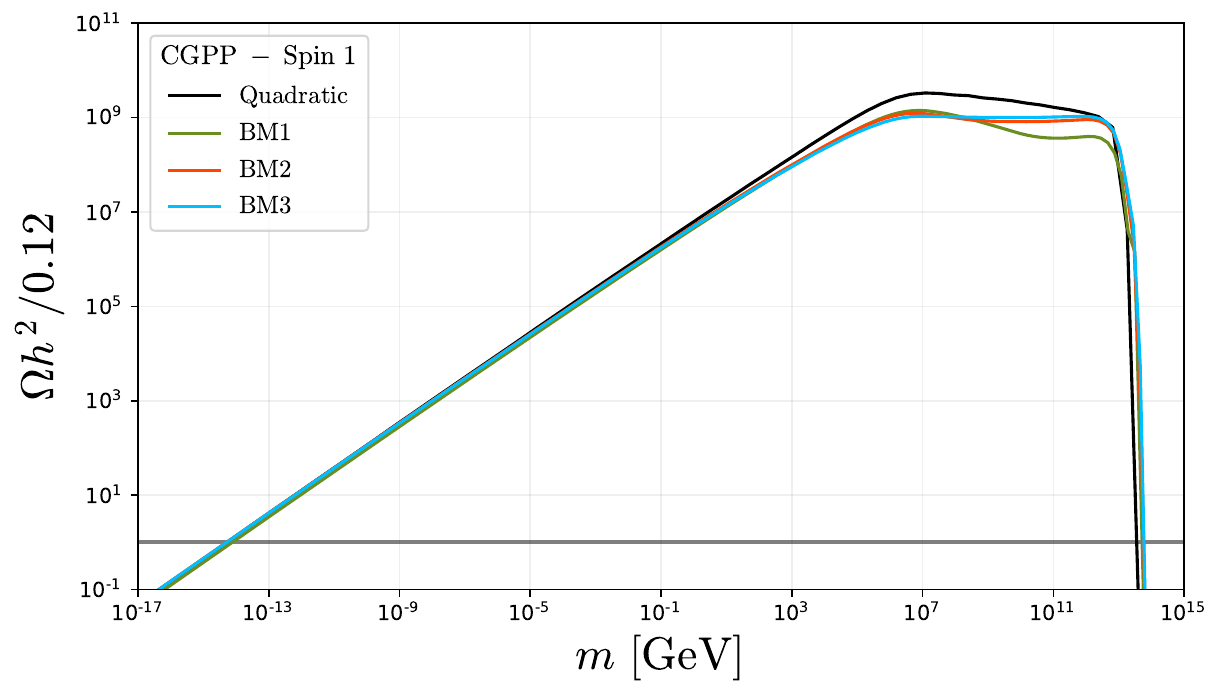}
    \caption{Abundance of spin-1 particles produced via CGPP as a function of the particle's mass. In black we show the result for the quadratic inflationary model of Eq.\,\eqref{eq:chaotic_potential}, while the green, red and blue curves are the results using the potentials in Eq.\,\eqref{eq:inflaton_potential_PBHs} for the benchmark values of Tab.\,\ref{tab:exam_pbh}. The gray horizontal line denotes $\Omega h^2/0.12 = 1$. Here, PBH evaporation is not taken into account. We use $H_e=10^{13}$ GeV and $T_\text{RH} =10^{12}$ GeV.}
    \label{fig:modified_potential}
\end{figure}
%%%%%%%%%%%%%%%%%%%%%%%%%%%%%%%%%%%%%%%%%%%%%%%%
The main results of our study are shown in Figs.\,\ref{fig:evol_USR}-\ref{fig:abundance}. To be systematic in our discussion, we will first discuss how the inflationary potentials of Eq.\,\eqref{eq:inflaton_potential_PBHs} affect the abundance of particles produced via CGPP. We then investigate how the PBHs impact CGPP of spin-1 and spin-0 bosons. Lastly, we compare the abundance of DM particles produced from CGPP in the presence of the PBHs, and from the PBHs themselves.

%%%%%%%%%%%%%%%%%%%%%%%%%%%%%%%%%%%%%%%%%%%%%%%%
\subsection{Effect on the CGPP abundance of the inflationary potential}
%%%%%%%%%%%%%%%%%%%%%%%%%%%%%%%%%%%%%%%%%%%%%%%%

The first question we ask ourselves is: {\it do the inflationary potentials that give rise to PBH production in the early universe affect in a significant way CGPP?} 
As mentioned in Sec.\,\ref{sec:PBHs}, the ``speed breaker'' features of the potentials we consider cannot be too close to the minimum of the potential in order to avoid production of too light PBHs. On the other hand, as discussed in Sec.\,\ref{sec:GPP}, for most of the modes CGPP starts to be effective towards the end of inflation or later, in such a way that we expect the final DM abundance to be only mildly affected by the specific choice of potential that generate PBHs. 

To confirm our intuition, we select three different benchmark values for the ultra-slow-roll potentials, given in Tab.~\ref{tab:exam_pbh}, which lead to the background and spin-1 modes evolution for the same benchmark values as presented in Fig.~\ref{fig:evol_USR}.
In Fig.\,\ref{fig:modified_potential}, we show the final DM abundance against the DM mass for different choices of inflationary potentials: the quadratic potential of Eq.\,\eqref{eq:chaotic_potential} and the ultra-slow-roll potentials of Eq.\,\eqref{eq:inflaton_potential_PBHs}.
For simplicity we only show our results for a spin-1 DM candidate, but analogous results apply for spin-0. 
We stress that the abundance shown in Fig.\,\ref{fig:modified_potential} refers only to the one produced by CGPP, ignoring completely the effect of the PBHs evaporation, that will be added later. 
As can be seen, the effect of changing the potential is rather mild and affects primarily the ``plateau'' part of the curve. On the contrary, the regions in which the abundance curve intersects the $\Omega h^2 \simeq 0.12$ line are basically unaffected by the choice of the potential. The numerical results shown in Fig.\,\ref{fig:modified_potential} confirm that the bulk of CGPP happens towards the end of inflation or later, where the potentials of Eq.\,\eqref{eq:inflaton_potential_PBHs} amount to small deformations of the quadratic potential of Eq.\,\eqref{eq:chaotic_potential}. 

For the previous reasons, in what follows we will always use Eq.\,\eqref{eq:chaotic_potential} for our numerical computations, and disregard the origin of the PBH population.
We take $H_e = 10^{13}$ GeV and $T_\text{RH} = 10^{12}$ GeV, and, for the sake of clarity in our analysis, we adopt the assumption that the PBH population follows a monochromatic mass distribution. 
We also assume that the formation temperature equals the reheating temperature, so that the PBH mass is determined according to the relation in Eq.\,\eqref{eq:temp_BHMass}. For $T_\text{RH}=10^{12}$ GeV, this fixes the PBH mass to $M_\text{BH}=1.8\times 10^7$ g. Furthermore, we treat the initial fraction $\beta$ as an independent parameter in our model. Note that, based on Eq.\,\eqref{eq:beta_crit}, a phase of PBH dominance takes place if $\beta\gtrsim 10^{-12}$.

%%%%%%%%%%%%%%%%%%%%%%%%%%%%%%%%%%%%%%%%%%%%%%%%%%
\subsection{Effect of a PBH population on the final DM abundance}
%%%%%%%%%%%%%%%%%%%%%%%%%%%%%%%%%%%%%%%%%%%%%%%%%%

We now turn to the main question of this paper: {\it how does the presence of a PBH population affect the final boson DM abundance?} We answer this question in three stages: first, we compute the spectrum of gravitationally produced DM, $n_k/(a_e H_e)^3$, as a function of the comoving momentum $k$, where $\,n_k \equiv k^3 |\beta_k|^2/(2\pi^2)$ 
, turning on the PBHs but {\it without} considering the DM population produced by the PBHs evaporation (see Fig.\,\ref{fig:spectra}). We then compute the total DM abundance of {\it gravitationally produced} bosons in the presence of a PBH population (i.e. considering the PBH domination phase and the entropy injection due to PBH evaporation, but without considering the DM population produced during evaporation) as a function of the boson mass (see Fig.\,\ref{fig:abundance}, left panels). Finally, we compute the total DM abundance considering both the contribution from CGPP and PBH evaporation (see Fig.\,\ref{fig:abundance}, right panels).

\subsubsection{Spectrum}

Regarding the spectrum, we consider a minimally coupled scalar ($\xi=0$) with mass $m = 10^{-30}$ GeV and a spin-1 particle of mass $m = 10^{-12}$ GeV. 
We have chosen these values of masses because they present all possible phenomenological effects regarding the interplay between CGPP and PBHs. Without PBHs, the spectrum for the minimally coupled spin-0 particle is expected to be constant for small values of $k$ and then to decrease as $k^{-1}$ once $k \gtrsim k_\star$ (the comoving momentum $k_\star$ is defined via $k_\star = a_\star m$, where $a_\star$ is such that $H(a_\star) = m$ and represents the moment in which the comoving Compton wavelenght equals the comoving horizon). While this behaviour is not intuitive, it is known in the literature\,\cite{Ema:2015dka,Ema:2016hlw,Chung:2018ayg,Chung:1998bt,Hashiba:2021npn,Li:2019ves}. For spin-1 particles, again without PBHs, the spectrum starts growing as $k^2$ for small values of $k$, and decreases as $k^{-1}$ for $k\gtrsim k_\star$\,\cite{Graham:2015rva,Ahmed:2020fhc,Kolb:2020fwh}.

\begin{figure}[t!]
\centering
    \includegraphics[width=0.48\linewidth]{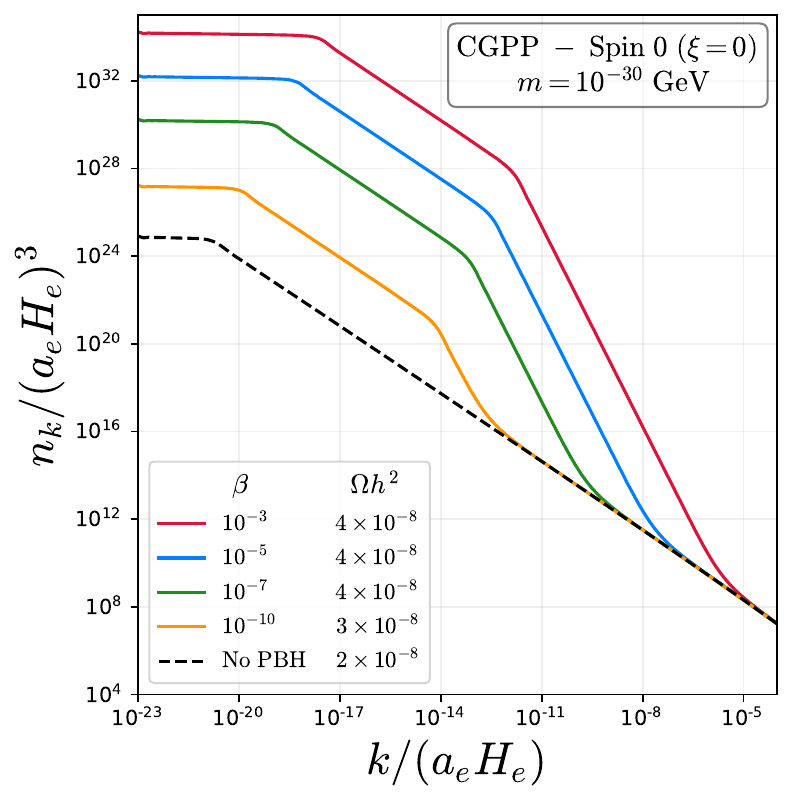}\includegraphics[width=0.443\linewidth]{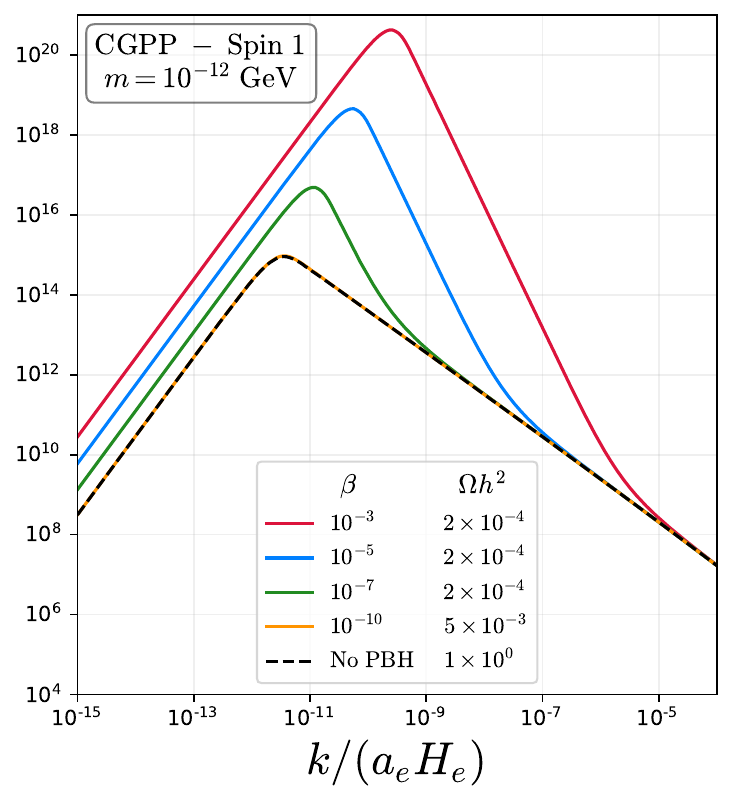}
    \caption{Dimensionless momentum spectrum $n_k/(a_eH_e)^3$ of particle production via CGPP in the presence of PBH population as a function of the dimensionless comoving momentum $k/(a_eH_e)$. \textbf{Left panel:} 
    Spectra for a minimally coupled scalar DM particle with mass $m=10^{-30}$ GeV for several values of the PBH abundance fraction $\beta$. \textbf{Right panel:} Same as left panel for a vector DM particle with mass $m=10^{-12}$ GeV. For both plots, we also highlight the corresponding values of abundance produced from each spectrum. We have used $H_e=10^{13}$ GeV, $T_\text{RH}=10^{12}$ GeV and $M_\text{BH}=1.8\times 10^7$ g.}
    \label{fig:spectra}
\end{figure}

What happens once a PBH population is present? We present in Fig.\,\ref{fig:spectra} the spectra $n_k$ as a function of $k$, where curves of different colors correspond to different choices of $\beta$ (and, as a consequence, of different values for $\Omega h^2$, stressing that this latter accounts for the contribution coming from CGPP alone). On the left panel we show the results for the minimally coupled spin-0 case ($\xi=0$), while on the right panel we show the spin-1 case. For the conformally coupled scalar with $\xi = 1/6$, the behaviour is similar to the latter and we do not show it explicitly. 

%%%%%%%%%%%%%%%%%%%%%%%%%%%%%%%%%%%%%%%%%%%%%%%%
%%%%%%%%%%%%%%%%%%%%%%%%%%%%%%%%%%%%%%%%%%%%%%%%
The behaviour of the curves, although similar to the one described above, presents a richer structure. As expected, the spectrum of the minimally coupled spin-0 DM is approximately constant for very low momenta, $n_k\sim k^0$. For $k > k_\star$, the spectrum begins to decrease, first as $k^{-1}$, then as $k^{-3}$ and then again as $k^{-1}$. The decrease as $k^{-3}$, not present in the spectrum without PBHs, is due to the phase of matter domination generated by the PBH population. For the vector DM instead, the spectrum starts growing as $k^2$, reaching its peak around $k_\star$, and then falls first as $k^{-3}$ and later as $k^{-1}$. Again, the former corresponds to the period of PBH domination. In both cases, though the shape of the spectra are very distinct, we observe that the presence of the PBHs has two common effects: ($i$) it increases the absolute value of the spectrum for momenta below a certain threshold and ($ii$) it increases the value of $k_\star$, thus increasing the values of $k$ at which the peak occurs in the spin-1 case and where the plateau of the spin-0 spectrum ends. 

Both effects are explained by the additional matter domination phase due to the PBH population which, for the values of the mass considered, takes place before $H=m$. On the one hand, this phase increases the number of gravitationally produced DM particles, hence increasing the absolute value of the spectrum. On the other hand, this phase slows down the evolution of the comoving horizon $1/aH$, delaying the moment at which $H=m$ and hence increasing the value of $k_\star = a_\star m$. Another interesting feature that emerges in Fig.\,\ref{fig:spectra} is that, for $k \gg k_\star$, the curves converge and the effect of the presence of the PBH population becomes irrelevant. This is due to the fact that, in this region of comoving momentum, CGPP concludes before the onset of the PBH domination phase, which is thus irrelevant for the computation of the number of gravitationally produced relic particles. 

It is worth stressing that, even though the spectrum might increase because of the PBHs, the resulting abundance is not necessarily larger. In the minimally coupled spin-0 scenario depicted in Fig.\,\ref{fig:spectra}, we see that the abundance is essentially unchanged by varying $\beta$, while for spin-1 the value of $\Omega h^2$ actually decreases when $\beta$ is increased. More in detail, we see in the right panel that the spectra for $\beta=10^{-10}$ and without PBHs are equal, but the abundances differ by orders of magnitude. In general, as we increase the DM mass, the spectra for different values of $\beta$ start to coincide with the one without any PBHs, whereas the corresponding $\Omega h^2$ decrease the larger $\beta$ is. As we will see shortly, this is a consequence of the entropy dilution due to PBH evaporation, which shows how non-trivial the interplay between the PBH dynamics and CGPP is.

\subsubsection{Abundance from CGPP}\label{subsec:CGPP_abundance}

\begin{figure}[t!]
\centering
    \includegraphics[width=0.54\linewidth]{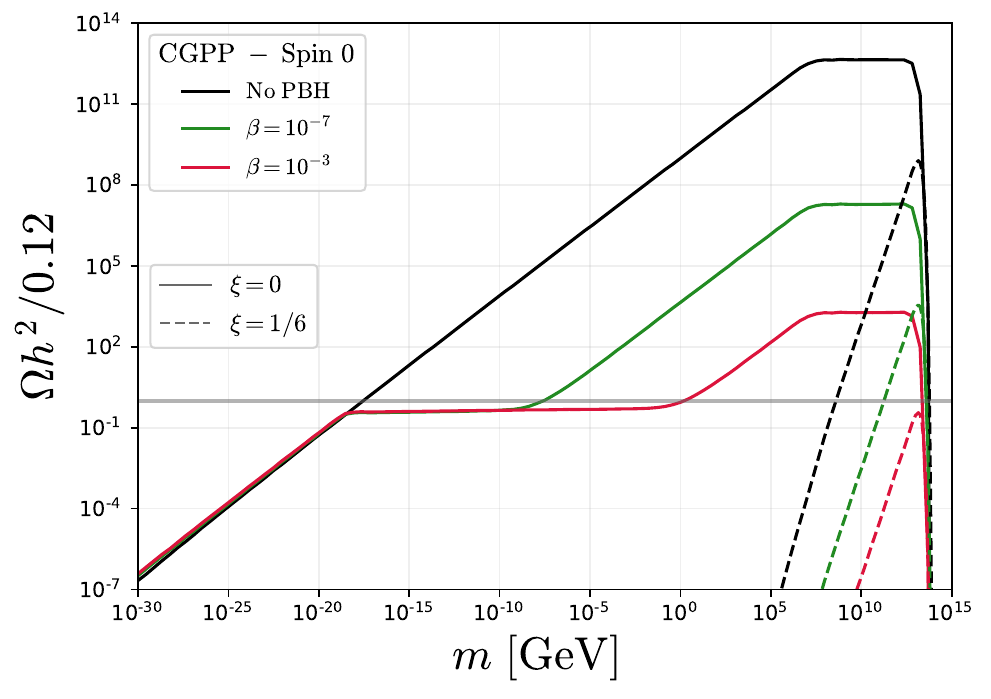}
    \includegraphics[width=0.43\linewidth]{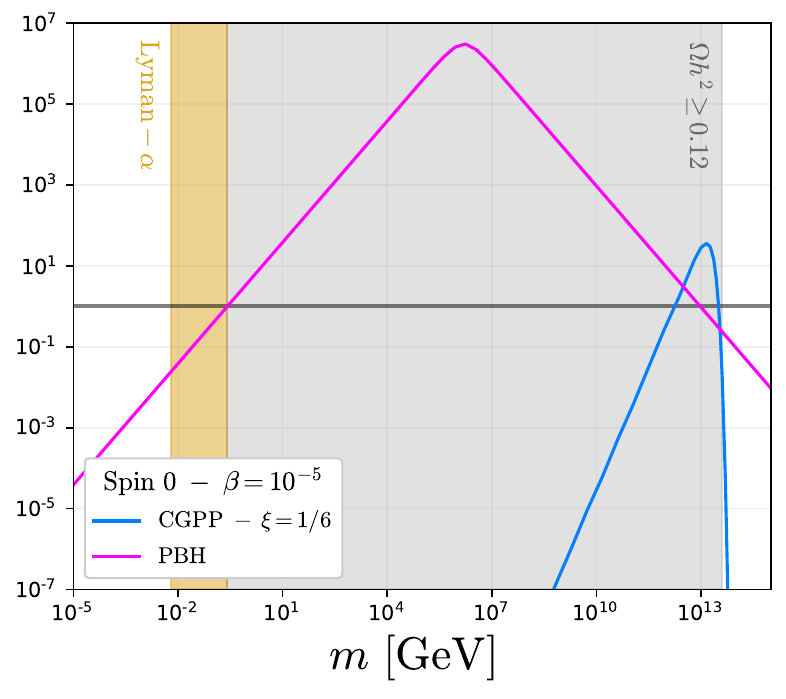}
    \includegraphics[width=0.54\linewidth]{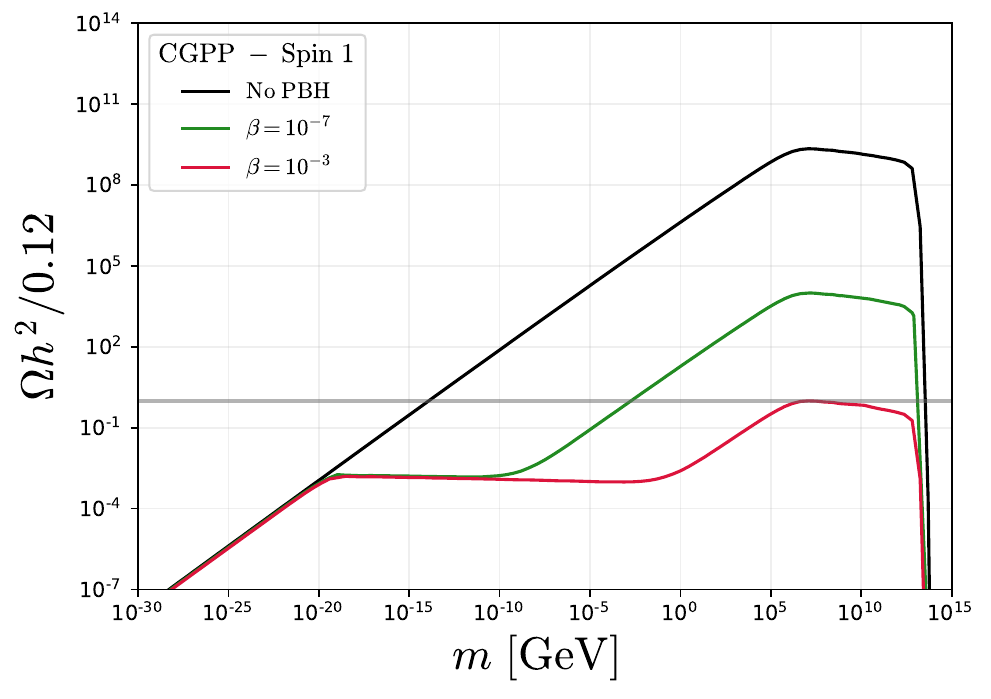}
    \includegraphics[width=0.43\linewidth]{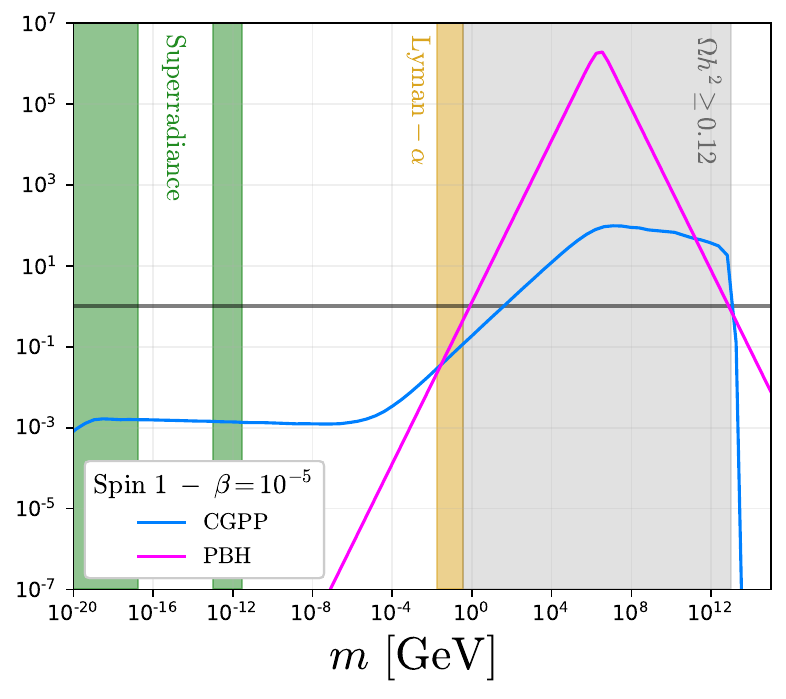}
    \caption{
    Abundance of particles produced by CGPP and PBHs. \textbf{Left Panels:} The abundance of DM coming from CGPP for a minimally and conformally coupled scalar (top panel) and for a vector (bottom panel) as a function of the DM mass. Black lines correspond to not having any PBHs, while for green and red ones we assume a PBH population with $\beta=10^{-7},~10^{-3}$, respectively. \textbf{Right panel:} Contributions to the abundance of a bosonic DM particle from CGPP (blue) and from PBH evaporation only (pink), assuming $\beta=10^{-5}$, as a function of the particle's mass $m$. The top (bottom) panel is for a conformally coupled spin-0 (spin-1) DM candidate. We show constraints from overabundance (gray), Lyman-$\alpha$ (yellow) and super-radiance (green). In both plots we use $H_e=10^{13}$ GeV, $T_\text{RH}=10^{12}$ GeV and $M_\text{BH}=1.8\times 10^7~\text{g}$.}
    \label{fig:abundance}
\end{figure}

We now turn to Fig.\,\ref{fig:abundance}. On the left panels, we show the DM abundance for a spin-0 DM candidate with $\xi = 0$ (solid) and $\xi = 1/6$ (dashed) and for a spin-1 candidate as a function of the DM mass {\it without} considering the DM population produced by the PBH evaporation but including the entropy injection due to evaporation. Black lines apply when no PBHs population is considered, while the red (green) lines apply when a PBHs population is present with $M_{\rm BH} = 1.8\times 10^7$ g and $\beta = 10^{-3}~(10^{-7})$.

Let us start with the black lines, when there is no PBH population. 
The behaviour of the curves for a minimally coupled spin-0 DM candidate and for a spin-1 candidate is similar: $\Omega h^2$ grows as $m^{1/2}$ for small $m$, reaches a plateau and then abruptly decreases for $m \gtrsim H_e$\,\cite{Graham:2015rva,Ema:2019yrd,Ahmed:2020fhc,Kolb:2020fwh,Kolb:2023ydq}. For the masses in the plateau, the final abundance (i.e. the moment in which $H = m$) happens during reheating, while for masses on the left of the plateau it happens during radiation domination. 

Adding now a PBH population, we see two main effects: ($i$) the absolute value of $\Omega h^2$ decreases and ($ii$) a new plateau appears at smaller $m$. The first effect is easily explained by the huge entropy injection because of BH evaporation (see Eq.\,\eqref{eq:entropy_dilution}). Since this happens after CGPP ceases to be effective (i.e. after $H=m$), the net effect is to diminish the final DM abundance. The second effect is instead due to the phase of matter domination owing to the PBH population: the additional plateau appears for those masses for which $H=m$ happens during this phase, and disappears for smaller masses for which it is still true that $H=m$ during the radiation domination phase that follows the PBH domination. This explains why for these masses the curves with or without PBHs coincide: $H=m$ happens so late that any PBH dynamics ``decouples'' from the determination of the DM abundance. In particular, also the entropy dilution is ineffective, since CGPP finishes well after the PBH population has completely evaporated, in such a way that the final DM abundance is not affected by this change in the SM bath temperature. 

For a spin-0 DM with $\xi = 1/6$ the dependence on $m$ is much steeper than in the cases discussed above and the effect of matter domination phases much less pronounced. This is due to the different mode functions that enter in the computation of the CGPP abundance (see App.\,\ref{app:approximate_solutions_scalar} and Fig.\,\ref{fig:conformal_scalar_Oh2} for details). 

In Fig.\,\ref{fig:abundance} we have shown the DM abundance for a definite choice of $H_e$ and $T_\text{RH}$ (or, equivalently, of $M_\text{BH}$). It is easy to understand how $\Omega h^2$ would change for other choices of these parameters. By changing $H_e$, the end point of the curves in the left panels of Fig.\,\ref{fig:abundance} is shifted, since for $m\gtrsim H_e$ CGPP is exponentially suppressed. Also, the curves are dislocated up and down according to Eq.\,\eqref{eq:OmegaDM_approx}. By increasing (decreasing) the value of $T_\text{RH}$, we obtain a smaller (larger) value of $M_\text{BH}$, as we assume them to be related by Eq.\,\eqref{eq:temp_BHMass}. If we lower the value of $M_\text{BH}$ in Fig.\,\ref{fig:abundance}, the plateau corresponding to the phase of PBH dominance is shifted up and is reduced in size, because the extra phase of matter dominance is shorter and the PBHs evaporate faster. The exact same effect happens also for the plateau of the reheating phase, since larger $T_\text{RH}$ implies a shorter period of reheating.

%%%%%%%%%%%%%%%%%%%%%%%%%%%%%%%%%%%%%%%%%%%%%%%%
%%%%%%%%%%%%%%%%%%%%%%%%%%%%%%%%%%%%%%%%%%%%%%%%

\subsubsection{Total abundance and constraints}

Finally, we present the total DM abundance, showing the contributions from both CGPP and PBH evaporation, on the right panels of Fig.\,\ref{fig:abundance}, fixing $\beta = 10^{-5}$. The blue curves reproduce the abundance obtained considering only CGPP (as in the left panels of the same figure), while the pink curves represent the DM abundance produced only by the PBH evaporation. Clearly, the total abundance is obtained summing the two contributions. Upper and lower panels show the results for spin-0 and spin-1, respectively.

The behavior of the DM abundance produced by the PBH evaporation can be understood as follows: the peak corresponds to a value of the DM mass equal to the initial BH temperature ($T_\text{BH} \simeq 5 \times 10^5$ GeV for the BH mass we are considering, see Eq.\,\eqref{eq:T_BH}). For smaller masses, the abundance grows as $m$, while for larger values it decreases as $m^{-1}$\,\cite{Gondolo:2020uqv}. As can be seen from the plots, the presence of a DM population produced by the PBH evaporation can drastically change the total abundance. For spin-1 DM (lower panel), the effect is pretty large and the evaporation population dominates the abundance for $10^{-1}\,{\rm GeV} \lesssim m \lesssim 10^{11}\,{\rm GeV}$. A similar conclusion is true for a conformally coupled spin-0 DM candidate (upper panel), for which the population produced by evaporation dominates the abundance for most of the parameter space. We do not show the abundance for the minimally coupled spin-0 candidate, since this case is almost entirely excluded by isocurvature constraints\,\cite{Chung:2004nh,Chung:2011xd,Garcia:2023qab}. Clearly, the position peak of the DM population produced by evaporation depends strongly on the chosen value of the BH temperature. For smaller (larger) values of $T_{\rm BH}$, the peak would move to the left (right), changing the region in which evaporation dominates the DM abundance. Regarding the abundance of CGPP particles, the dependence of the curve on the parameters $H_e$ and $T_\text{RH}$ has been discussed in detail at the end of Sec.\,\ref{subsec:CGPP_abundance}.

In addition to the DM abundances from CGPP and from PBH evaporation, we highlight on the right panels of Fig.\,\ref{fig:abundance} the regions which are disfavored by experimental observations. In gray, we have the region where DM is overabundant. 
In yellow, we show an estimate of the constraints coming from Lyman-$\alpha$ data\,\cite{Hui:1996fh,Gnedin:2001wg}, which can be translated to a bound on the fraction of warm DM. 
Although a more precise determination of the bound requires the determination of the time evolution of the DM phase space, for our purposes, applying a simple criterion will suffice to determine whether the DM is too warm. Since, for our choice of parameters, the abundance close to the observed value is dominated by the evaporation process, we specialize our discussion to this case. In any case, the argument can easily be adapted to cases in which the abundance close to the measured value is dominated by CGPP, a situation which may happen for other choices of parameters.\footnote{Production of warm DM from CGPP is typically relevant for masses $m\lesssim 1$ eV. In the spin-0 case, see for instance Ref.\,\cite{Garcia:2023qab}.} 
We require that the average DM velocity today, obtained from the average momentum at emission, be lower than the limit velocity of warm DM from Lyman-$\alpha$\,\cite{Baldes:2020nuv,Masina:2021zpu, Cheek:2021odj}
\begin{align}
    v_{\rm DM} = \frac{a_{\rm ev}}{a_0}\frac{\langle p_{\rm DM} \rangle}{m} \lesssim 3.9\times 10^{-8} \left(\frac{1~{\rm keV}}{m}\right)^{4/3},
\end{align}
with $a_{\rm ev}$, $a_0$ the scale factors at evaporation and today, respectively.
Using the results of Refs.\,\cite{Boyarsky:2008xj,Baur:2017stq}, we have that for DM masses smaller that 1 keV, the fraction of the total DM that is warm must be less that about 2\%. For warm DM emitted from PBHs, we can use the results from Refs.\,\cite{Auffinger:2020afu,Baldes:2020nuv,Cheek:2021odj,Cheek:2021cfe}, which exclude DM masses below around 1 GeV. 
Thus, we impose the conservative bound of $(\Omega h^2/0.12)_\text{warm}\lesssim 0.02$ for $m\lesssim 1$ GeV. 
Lastly, in green, we show the bounds from BH super-radiance\,\cite{Arvanitaki:2010sy,Brito:2015oca}. 
This phenomenon consists in the depletion of the spin of BHs of astrophysical nature due to the exponential growth of the number occupation of bosons gravitationally bounded to the BH.
On the top right panel of Fig.\,\ref{fig:abundance}, BH super-radiance of spin-1 particles constraint the window $10^{-13}\lesssim m/\text{GeV}\lesssim 3\times 10^{-12}$\,\cite{Cardoso:2018tly}, while more massive BHs exclude the region $10^{-22}\lesssim m/\text{GeV}\lesssim 10^{-17}$\,\cite{Unal:2020jiy,Chen:2022nbb,Saha:2022hcd}.

\subsection{Production of PBHs from CGPP}

Up to this point, we have always considered a PBHs population generated during inflation by some peak in the power spectrum, see App.\,\ref{app:power_spectrum}. 
However, as can be seen from Fig.\,\ref{fig:spectra}, also the spectrum of gravitationally produced bosonic DM is peaked, with a corresponding peak appearing also in the power spectrum. This motivates the question: \textit{can gravitationally produced DM generate a PBH population?}

For spin-1 DM, the answer seems to be negative. As shown in Ref.\,\cite{Gorghetto:2022sue}, the peak of the spectrum falls into a region in which the quantum pressure is important. This implies that an important fraction of DM is found in the form of complex structures, i.e. solitons. This is due to the fact that quantum pressure is important for modes $k \gtrsim k_J \equiv a(16 \pi G \rho m^2)^{1/4}$ (where $G$ is Newton's constant) and, as shown in Ref.\,\cite{Gorghetto:2022sue}, the peak in the spin-1 DM spectrum happens at values $k_\star \sim k_J/2$, i.e. not too far from the region in which quantum pressure dominates.

Turning to spin-0, we see from Fig.\,\ref{fig:spectra} that no peak appears for the minimally coupled case. However, a peak similar to the spin-1 case appears for a conformally coupled scalar, although we do not show explicitly this case in Fig.\,\ref{fig:spectra}. This means that, in principle, also in this case we can have PBH production, provided the peak of the spectrum falls in a region in which quantum pressure is not important. To estimate if this is the case, we can proceed as follows. Let us assume instantaneous reheating. Then the peak of the spectrum is approximately at $k_\star = a_\star m$, where $a_\star$ is defined by the condition $H(a_\star) = m$ (see App.\,\ref{app:approximate_solutions_scalar} for analytic approximations). Computing the ratio $k_\star/k_J$ at $a_\star$, we obtain $k_\star/k_J = \sqrt{m_{PL} m}/(2 m [na^3]/a_\star^3)^{1/4}$, where we introduced the reduced Planck mass $m_{PL}$, we used the definition $k_J =a(16 \pi G \rho m^2)^{1/4} $\,\cite{Gorghetto:2022sue} and we wrote the DM density as $\rho = m n = m [na^3]/a_\star^3$ using the fact that, after we reach $a_\star$ when $H=m$, the comoving density $[na^3] \simeq$ const. We can now use the approximate analytic expression for $na^3$ obtained in Eq.\,\eqref{eq:na3_analytical_Rad} of App.\,\ref{app:approximate_solutions_scalar} (taking $a_\text{RH} = a_e$ and $H_\text{RH} = H_e$ since we are considering instantaneous reheating) to estimate $k_\star/k_J \sim \sqrt{m_{PL}/m} \gg 1$, where the expression is valid apart from $\mathcal{O}(1)$ numbers. Our estimate thus indicates that, also in this case, the peak of the spectrum falls in the region in which quantum pressure is important.
It thus seems that also in the case of spin-0 DM no PBH formation happens. Since, however, a dedicated study would be needed to assess whether our conclusions are unavoidable or some way out can be found, we defer such analysis to future work.

%%%%%%%%%%%%%%%%%%%%%%%%%%%%%%%%%%%%%%%%%%%%%%%%%%%%%%%%%%%%%%%%%%%%%%%%%%%%%%%%%%%%%%%%%%%%%%%%%%%%%%%%%%%%%%%%%%%%%%%%%%%%%%%%%%%%
\section{Conclusions}\label{sec:conclusions}
%%%%%%%%%%%%%%%%%%%%%%%%%%%%%%%%%%%%%%%%%%%%%%%%%%%%%%%%%%%%%%%%%%%%%%%%%%%%%%%%%%%%%%%%%%%%%%%%%%%%%%%%%%%%%%%%%%%%%%%%%%%%%%%%%%%%

In this work we have analysed the interplay between two phenomena that may take place in the early universe: gravitational production of bosonic DM (more specifically, spin-0 and spin-1) and the existence of a PBH population. The question of what happens to the DM abundance in the presence of PBHs is not new and has been analysed in a number of scenarios, for example when PBH physics acts in the presence of a DM population produced via freeze-out or freeze-in. In these cases, the effects of the PBHs are two-fold: ($i$) BH evaporation produces an additional DM population and ($ii$) the huge entropy injection due to BH evaporation tends to dilute the abundance of DM created via other mechanisms. In the case under consideration, in which DM is generated via CGPP in the early universe, the interplay is more subtle: in addition to the two effects just mentioned, a possible early matter domination phase due to PBH dominance causes important qualitative and quantitative differences on the final DM spectrum and abundance, provided this phase happens before gravitational production has ended. 
This can be clearly seen in Fig.\,\ref{fig:spectra}, where the DM spectrum of gravitationally produced DM is shown, and in Fig.\,\ref{fig:abundance}, where we present the final DM abundance, with and without the contribution from PBH evaporation. We thus conclude that, unlike what happens, for instance, in the freeze-out and freeze-in cases mentioned above, the presence of a PBH population cannot be disentangled from CGPP, because the final number of gravitationally produced DM particles depends on the PBH abundance and mass. The computation of the total DM abundance can thus be performed only once all the parameters of the theory are fixed, both in the CGPP sector ($m$, $H_e$, $T_{\rm RH}$) and in the PBH sector ($M_{\rm PBH}$, $\beta$). Depending on the choice of parameters, the correct abundance can be dominated by either the evaporation or the gravitationally produced population.

Our study can be extended in a number of directions. First of all, we have considered the simplest case in which the PBHs do not have spin and have a monochromatic population. It would be interesting to study what happens relaxing these conditions and see how the PBH mass distribution and the spin are reflected into the CGPP. Moreover, we have considered a scenario in which PBHs are produced by gravitational collapse of the density fluctuations. 
Alternatives for the formation of PBHs, such as phase transitions \,\cite{Baker:2021nyl,Lewicki:2023ioy} or collapse due to additional Yukawa interactions\,\cite{Amendola:2017xhl,Savastano:2019zpr,Flores:2021jas,Flores:2024lng,Flores:2024eyy}, would lead to distinct predictions for the PBH parameters and possibly modify the cosmological history, and accordingly CGPP, in other ways. From the CGPP side, one could consider physical scenarios that lead to a different evolution of the universe. For example, in theories with axions or axion-like particles, an extra phase of kination may take place\,\cite{Gouttenoire:2021jhk,Co:2021lkc,Co:2020jtv,Co:2019wyp,Gouttenoire:2021wzu} and consequently have important effects on the final number of DM produced from CGPP. 
Similarly, heavy and long-lived degrees of freedom could have dominated the energy density before BBN, resulting in a different type of matter-dominated era that would not end as suddenly as the one caused by PBHs. In this scenario, we would also anticipate modifications to the CGPP.
We leave the study of these questions to future work.

\acknowledgments 

We would like to thank A.~Long for useful discussions on CGPP.
The work of EB is partly supported by the Italian INFN program on Theoretical Astroparticle
Physics (TAsP), by ``Funda\c{c}\~ao de Amparo \`a Pesquisa do Estado de S\~ao Paulo”
(FAPESP) under contract 2019/04837-9, as well as by Brazilian “Conselho
Nacional de Deselvolvimento Cient\'ifico e Tecnol\'ogico” (CNPq).
The work of YFPG has been funded by the UK Science and Technology Facilities Council (STFC) under grant ST/T001011/1. 
GMS acknowledges financial support from "Fundação de Amparo à Pesquisa do Estado de São Paulo" (FAPESP) under contracts 2020/14713-2 and 2022/07360-1. GMS is grateful to the Theory Group of the Deutsches Elektronen-Synchrotron (DESY) for the hospitality during early stages of this work.
RZF is also partly supported by FAPESP and by CNPq.
This project has received funding/support from the European Union’s Horizon 2020 research and innovation programme under the Marie Sk\l{}odowska-Curie grant agreement No 860881-HIDDeN.
This work has made use of the Hamilton HPC Service of Durham University.

\appendix

\section{CGPP of a Conformally coupled scalar: analytical and numerical computations}\label{app:approximate_solutions_scalar}

\subsection{Approximate analytical solutions}

In this Appendix we elaborate on the approximate analytical solutions to Eq.\,\eqref{eq:EoM_scalar} in the case $\xi = 1/6$ (analytical solutions were also discussed for instance in Refs.\,\cite{Ema:2015dka,Ema:2016hlw,Chung:2018ayg,Chung:1998bt,Hashiba:2021npn,Li:2019ves}). The differential equation is given by
\be\label{eq:EoM_scalar_app}
v_k''(\eta) + \omega_k^2(\eta)v_k^{\phantom{2}}(\eta)=0,\quad \omega_k^2(\eta) = k^2 + a(\eta)^2m^2,
\ee
which is not exactly solvable for general $k$ and $a(\eta)$.

Our first step is to approximate the cosmological evolution: we assume a perfect de Sitter space-time for inflation, followed by an instantaneous transition to reheating and afterwards to a radiation dominated epoch. In terms of conformal time, the scale factor for each period is given by
\be\label{eq:scale_factor_approximate}
a(\eta) = a_r\left[\frac{1+3\omega}{2}a_rH_r(\eta - \eta_r)+1\right]^{\frac{2}{1+3\omega}},
\ee
where $\omega = -1,\,0,\,1/3$ is the equation of state for inflation, reheating and radiation domination, respectively. Furthermore, $a_r,\,H_r,\,\eta_r$ are equal to $a_e,\,H_e,\eta_e$ for inflation and reheating, and to $a_\text{RH},\,H_\text{RH},\,\eta_\text{RH}$ for radiation domination. We remind that the subscript $e$ ($\text{RH}$) denotes quantities at the end of inflation (reheating).  During inflation, it follows from Eq.\,\eqref{eq:scale_factor_approximate} that the limit $a(\eta)\to 0$ corresponds to $\eta\to -\infty$.

The second approximation we make is to consider regions in the parameters that have simpler expressions for the frequency. More precisely, we introduce two regimes:
\al{
&\omega_k^2(\eta) \simeq k^2,\quad & k\gg a(\eta)m,\\ &\omega_k^2(\eta) \simeq  a(\eta)^2 m^2,\quad &k\ll a(\eta)m,
}
that correspond to different hierarchies between $k^2$ and $a(\eta)^2m^2$. This simplification will allow us to solve analytically Eq.\,\eqref{eq:EoM_scalar_app} and study the solutions in detail. We notice that for the first regime, $k\gg a(\eta)m$, when the modes are highly relativistic, the solution is independent of the scale-factor and given by a sum of positive and negative frequency modes,
\be\label{eq:k>>am_solution}
v_k(\eta) = b_+ e^{ik\eta } + b_- e^{-ik\eta},\quad k\gg a(\eta)m,
\ee
with $b_\pm$ constants. The evolution of all $k$-modes starts in a regime in which Eq.\,\eqref{eq:k>>am_solution} is valid, since for each $k$ there is a sufficiently early time for which $k\gg a(\eta) m$ is satisfied. The wave-function at very early times, $\eta\to -\infty$, is fixed by the Bunch--Davies initial condition\,\cite{Bunch:1978yq}:
\be
v_k^\text{BD} = \frac{e^{-ik\eta}}{\sqrt{2k}},
\ee
which therefore fixes $b_+=0$ and $b_-=1/\sqrt{2k}$ in Eq.\,\eqref{eq:k>>am_solution}.

\subsubsection{Inflation}

Let us now examine solutions during inflation in the non-relativistic regime, namely $k\ll a(\eta)m$. The equation of motion becomes
\be
v_k''(\eta) + \left[\frac{a_em}{1-a_eH_e\eta}\right]^2v_k^{\phantom{'}}(\eta) = 0,
\ee
where we have used Eq.\,\eqref{eq:scale_factor_approximate} with $\omega=-1$ for the scale factor. We have also set $\eta_e=0$. The solution reads
\be\label{eq:k<<am_inflation_nu}
v_k(\eta) = b_1\left(\frac{a(\eta)}{a_e}\right)^{-\frac{1-\nu}{2}} + b_2\left(\frac{a(\eta)}{a_e}\right)^{-\frac{1+\nu}{2}},\quad \nu^2 \equiv 1-\frac{4m^2}{H_e^2}.
\ee
Here, we can distinguish between two cases, $m/H_e\ll 1$ and $m/H_e\gg 1$, that correspond to $\nu\simeq 1$ and $\nu\simeq 2i m/H_e$, respectively.

Taking $m/H_e\ll 1$, the solution in Eq.\eqref{eq:k<<am_inflation_nu} becomes
\be
v_k(\eta)\simeq b_1 + b_2 \frac{a_e}{a(\eta)}.
\ee
The coefficients $b_{1,2}$ can be determined by joining $v_k$ and $v_k'$ to the ones from Eq.\,\eqref{eq:k>>am_solution} at $a_m = k/m$, value that sets the boundary between the $k \gg a(\eta) m$ regime in which Eq.\,\eqref{eq:k>>am_solution} applies and the $k \ll a(\eta) m$ regime in which Eq.\,\eqref{eq:k<<am_inflation_nu} is valid, resulting in
\be
v_k(\eta) \simeq \frac{e^{-ik\eta_m}}{\sqrt{2k}}\left[1-i\frac{m}{H_e} + \frac{ik}{a(\eta)H_e}\right],
\ee
with $\eta_m\equiv \eta(a_m)$. Hence, the solution for inflation is given by
\be\label{eq:solution_inflation}
v_k^\text{inf}(\eta) \simeq \left\{\begin{array}{ll}
\frac{e^{-ik\eta}}{\sqrt{2k}},\quad &  k\gg a(\eta) m,\\
\frac{e^{-ik\eta_m}}{\sqrt{2k}}\left[1-i\frac{m}{H_e} + \frac{ik}{a(\eta)H_e}\right],\qquad    & H_e \gg m \gg k/a(\eta).
\end{array}\right.
\ee
From the equation above, we can compute the corresponding Bogoliubov coefficients using Eq.\,\eqref{eq:Bogoliubov}. We obtain
\be\label{eq:Bogoliubov_inflation}
|\beta_k^\text{inf}|^2 \simeq \left\{\begin{array}{ll}
0,&\quad  k\gg a(\eta) m,\\
\frac{k}{4a(\eta)m}+ \frac{a(\eta)m}{4k}\left[1+\left(\frac{k}{a(\eta)H_e}-\frac{m}{H_e}\right)^2\right] - \frac{1}{2},    &\quad H_e \gg m \gg k/a(\eta).
\end{array}\right.
\ee
Notice that only after the mode exits the region in which $k\gg a(\eta)m$ particle creation begins to be effective.

If instead we consider $m\gg H_e$, that is, imaginary $\nu$, the solution in Eq.\,\eqref{eq:k<<am_inflation_nu} becomes highly oscillatory and the resulting $|\beta_k|^2$ vanishes on average. A more precise treatment\,\cite{Ema:2018ucl} shows that in this region CGPP is exponentially suppressed, so that we will not consider $m\gg H_e$ further in our analysis.

\subsubsection{Reheating}

During reheating, which we approximate as having equation of state $\omega =0$, we must solve the following differential equation when $k \ll a(\eta) m$:
\be
v_k''(\eta) + m^2 a_e^2 \left(1+\frac{a_eH_e}{2}\eta\right)^4v_k^{\phantom{'}}(\eta) = 0,
\ee
where we have again used that $\eta_e=0$. The equation above admits solutions in terms of Bessel functions $J_n$,
\be\label{eq:Bessel}
v_k(\eta) = c_1 z^{1/6}J_{-1/6}(z) + c_2 z^{1/6}J_{1/6}(z), \quad z\equiv \frac{2m}{3H},
\ee
with $c_{1,2}$ constants and $H$ the Hubble parameter. The solutions, and consequently the Bogoliubov coefficient, have very distinct behaviours depending on the value of $z$. More precisely, we can expand for different regimes of the variable $z$,
\be\label{eq:solution_RH_coeff}
v_k(\eta)\simeq \left\{\begin{array}{ll}
c_1' + c_2'\left[\frac{3H_e}{2m}\right]^{1/3} z^{1/3}, &\quad z\ll 1,\\
z^{-1/3}\left[c_1'' \cos(z)+c_2''\sin(z)\right],&\quad  z\gg 1.
\end{array}\right.
\ee
It is then easy to check that the corresponding Bogoliubov coefficient for $z\gg 1$ is simply a constant, while for $z\ll 1$ it has a non-trivial dependence on $a(\eta)$. This implies that particle production is only effective as long as $z \ll 1$, which can be translated to $m \ll H$.

We can determine the coefficients $c_{1,2}'$, as before, by matching the wave-function and its derivative to the solution during inflation\,\eqref{eq:solution_inflation} at $a(\eta_e)=a_e$. The matching procedure must now take into account different intervals in $k$, because the solution during inflation\,\eqref{eq:solution_inflation} has different expressions according to $k$. We identify three intervals: $[0,\bar k)$, $[\bar k, k_* )$ and $[k_*,\infty)$, where $\bar k \equiv a_e m$ and $k_* = a_* m$, with $a_*$ defined to satisfy $H(a_*)=m$.

In the first interval, $[0,\bar k)$, we have that $k\ll a(\eta)m$, meaning we must compare the second line from Eq.\,\eqref{eq:solution_inflation} with Eq.\,\eqref{eq:solution_RH_coeff} when $z\ll 1$. For the second interval, $[\bar k,k_*)$, we match the solution for $z\ll 1$ with the Bunch--Davies initial condition given in the first line of Eq.\,\eqref{eq:solution_inflation}. The wave-function for each case then becomes
\be\label{eq:solution_RH}
v_k^\text{RH}(\eta) \simeq\left\{\begin{array}{ll} 
\frac{e^{-ik\eta_m}}{\sqrt{2k}}\left[ 1+\frac{3ik}{a_eH_e}\left(1-\frac{a_em}{3k}\right) -\frac{2ik}{a_eH_e}\sqrt{\frac{a(\eta)}{a_e}} \right],&\quad k\in [0,\bar k),\\
\frac{e^{-ik\eta_m}}{\sqrt{2k}}\left[ 1+ \frac{2ik}{a_eH_e}\sqrt{\frac{k}{a_em}} - \frac{2ik}{a_eH_e} \sqrt{\frac{a(\eta)}{a_e}} \right],&\quad k\in [\bar k,k_*). 
\end{array}\right.
\ee
For modes in the last interval, $[k_*,\infty)$, consider first that $H=m$ is satisfied during reheating. Hence, the modes evolve from the Bunch--Davies vacuum directly to the solution with $z\gg 1$ in Eq.\,\eqref{eq:solution_RH_coeff}. Since the Bogoliubov coefficient produced by the latter is constant, and the one from the Bunch--Davies vacuum is zero, we find that no mode in this interval can produce particles. If $H=m$ is not achieved during reheating, we need to first obtain the solutions for radiation domination. However, we anticipate that the same conclusion will hold, i.e. we will still have that the Bogoliubov coefficient will be zero.

\begin{figure}[t!]
\centering
    \includegraphics[width=0.48\linewidth]{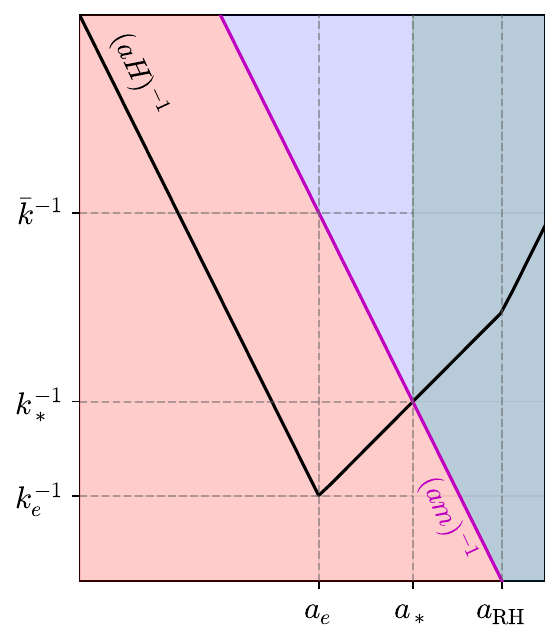}
    \includegraphics[width=0.495\linewidth]{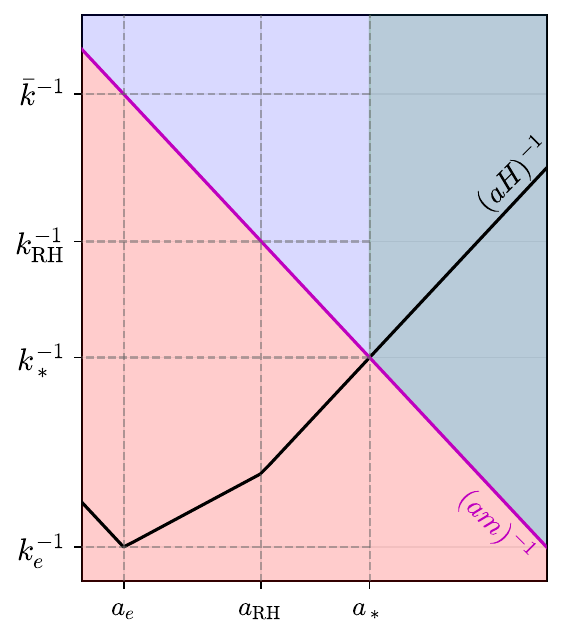}
    \caption{\textbf{Left panel:} Plot of the horizon $(aH)^{-1}$ (black curve) using the approximation of Eq.\,\eqref{eq:scale_factor_approximate}. In purple we show $(am)^{-1}$, that crosses the horizon during reheating. We make explicit in the vertical (horizontal) axis the values of comoving momenta (scale factor) that are relevant for the analytical computations. The regions depicted in red, lilac and green denote the ones in which the Bogoliubov coefficient is zero, when CGPP is efficient and when particle production ceases, respectively. \textbf{Right panel:} Same as left panel, but for $(aH)^{-1}=(am)^{-1}$ during radiation domination.}
    \label{fig:horizon_scalar}
\end{figure}

To help visualize the situation, we plot on the left panel of Fig.\,\ref{fig:horizon_scalar} the horizon $(aH)^{-1}$ in black and highlight the relevant scales, namely $\bar k$, $k_*$, $a_e$ and $a_*$. We choose $m$ such to have $m=H$ during reheating. The regime $k\gg a(\eta)m$ ($k\ll a(\eta)m)$ corresponds to the region below (above) the purple line. For $k\gg a(\eta)m$, denoted by the red region, the solution is given by Eq.\,\eqref{eq:k>>am_solution} and no particles are produced. In the lilac region particles are effectively produced, while in the green one, for which $H<m$, the Bogoliubov coefficient becomes constant.

%%%%%%%%%%%%%%%%%%%%%%%%%%%%%%%%%%%%%%%%%%%%%%%%
%%%%%%%%%%%%%%%%%%%%%%%%%%%%%%%%%%%%%%%%%%%%%%%%

\subsubsection{Radiation domination}

Lastly, consider the equation of motion\,\eqref{eq:EoM_scalar_app} with $\omega=1/3$ and $k\ll a(\eta)m$:
\be
v_k''(\eta) + m^2a_\text{RH}^2\left[1+a_\text{RH}H_\text{RH}(\eta - \eta_\text{RH})\right]^2v_k(\eta) = 0,
\ee
whose solution is
\be\label{eq:parabolic}
v_k(\eta) = d_1 D_{-1/2}((-1+i)x) + d_2 D_{-1/2}((1+i)x),\quad x\equiv \sqrt{\frac{m}{H}}
\ee
where $D_n$ is a parabolic cylinder function and $d_{1,2}$ are coefficients. Similar to the solution during reheating, the solution above is controlled by the ratio $m/H$ and we can expand for different regimes of $x$:
\be\label{eq:solution_rad_coeff}
v_k(\eta)\simeq \left\{\begin{array}{ll}
d_1' + d_2'\sqrt{\frac{H_\text{RH}}{m}} x, &\quad x\ll 1,\\
x^{-1/2}\left[d_1'' \cos(x^2/2)+d_2''\sin(x^2/2)\right],&\quad  x\gg 1.
\end{array}\right.
\ee
The oscillatory regime when $x\gg 1$ results in a constant Bogoliubov coefficient, while the solution for $x\ll 1$ allows for particle production.

In the same way we have analysed Eq.\,\eqref{eq:solution_RH_coeff} in terms of different intervals in momentum space, we identify four relevant intervals to match the solution during reheating to the one of Eq.\,\eqref{eq:solution_rad_coeff}: $[0,\bar k)$, $[\bar k,k_\text{RH})$, $[k_\text{RH},k_*)$ and $[k_*,\infty)$, where $k_\text{RH}=a_\text{RH}m$. As previously argued, wave-function with modes $k\in [k_*,\infty)$ will not produce particles effectively, since they evolve directly from the Bunch--Davies vacuum to a constant Bogoliubov coefficient. We report the solution for the remaining intervals:
\be\label{eq:solution_rad}
v_k^\text{rad}(\eta) \simeq\left\{\begin{array}{ll} 
\frac{e^{-ik\eta_m}}{\sqrt{2k}}\left[1 - \frac{2ik}{a_eH_e}\left(\sqrt{\frac{a_\text{RH}}{a_e}} -\frac{3}{2}\left(1-\frac{a_em}{3k}\right)-\frac{a_eH_e}{2a_\text{RH}H_\text{RH}}\right) - \frac{ik}{a_\text{RH}H_\text{RH}}\frac{a(\eta)}{a_\text{RH}}\right],\\ 
\qquad\qquad\qquad\qquad\qquad\qquad\qquad\qquad\qquad\qquad\qquad\qquad\qquad k\in [0,\bar k),\\
\frac{e^{-ik\eta_m}}{\sqrt{2k}}\left[1+\frac{2ik}{a_eH_e}\left(\sqrt{\frac{k}{a_em}}-\sqrt{\frac{a_\text{RH}}{a_e}}+\frac{a_eH_e}{2a_\text{RH}H_\text{RH}}\right)-\frac{ik}{a_\text{RH}H_\text{RH}}\frac{a(\eta)}{a_\text{RH}}\right],\\ 
\qquad\qquad\qquad\qquad\qquad\qquad\qquad\qquad\qquad\qquad\qquad\qquad\quad k\in [\bar k,k_\text{RH}),\\
\frac{e^{-ik\eta_m}}{\sqrt{2k}}\left[1+\frac{ik^2}{a_\text{RH}^2H_\text{RH}m}-\frac{ik}{a_\text{RH}H_\text{RH}}\frac{a(\eta)}{a_\text{RH}}\right],\quad k\in [k_\text{RH},k_*). 
\end{array}\right.
\ee
We do not report the expressions for the Bogoliubov coefficient corresponding to the different solutions just encountered since they are rather long and not particularly illuminating, but they can be easily computed using Eq.\,\eqref{eq:Bogoliubov}.

We show on the right panel of Fig.\,\eqref{fig:horizon_scalar} the horizon $(aH)^{-1}$, similar to the left panel, but with $m=H$ satisfied during radiation domination. We indicate the relevant momentum scales, $\bar k$, $k_\text{RH}$ and $k_*$. The red region has vanishing Bogoliubov coefficient, in the lilac one particle production is efficient and in the green region it ceases and the Bogoliubov coefficient becomes constant.

\subsubsection{Expressions for $na^3$}
%%%%%%%%%%%%%%%%%%%%%%%%%%%%%%%%%%%%%%%%%%%%%%%%
\begin{figure}[t!]
\centering
    \includegraphics[width=0.85\linewidth]{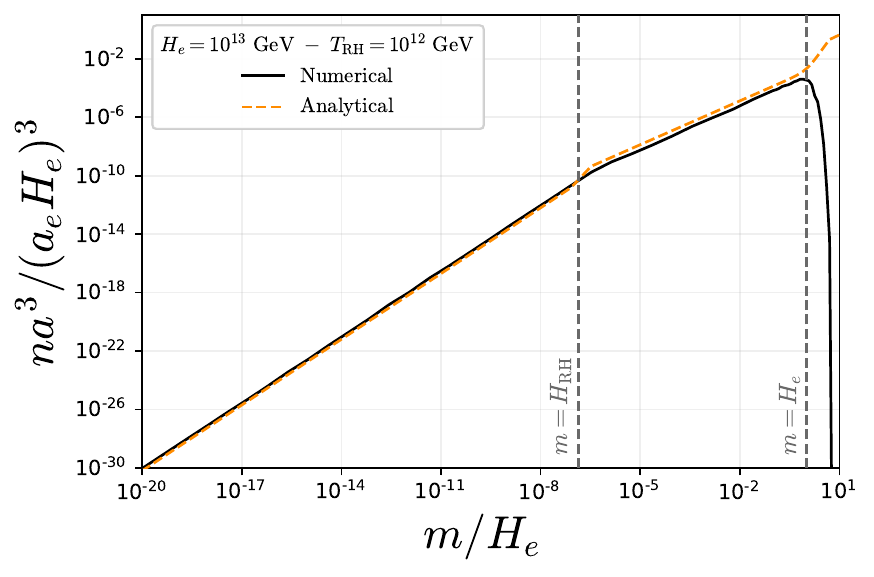}
    
    \caption{Results for the normalized comoving number density $na^3/(a_e H_e)^3$ of a conformally coupled scalar ($\xi = 1/6$) as a function of $m/H_e$ in the absence of PBHs. The black curve is obtained by numerically solving Eq.\,\eqref{eq:EoM_scalar_app} and in orange we show Eqs.\,\eqref{eq:na3_analytical_RH} and \eqref{eq:na3_analytical_Rad}. We have used $H_e=10^{13}$ GeV and $T_\text{RH}=10^{12}$ GeV, and we highlight with the vertical dashed lines the values of masses for which $m=H_e$ and $m=H_\text{RH}$.}
    \label{fig:scalar_analytical}
\end{figure}
%%%%%%%%%%%%%%%%%%%%%%%%%%%%%%%%%%%%%%%%%%%%%%%%
With the results in Eqs.\eqref{eq:solution_RH} and \eqref{eq:solution_rad} we can now compute the number density of particles produced via CGPP. As we have seen explicitly from the solutions in Eqs.\,\eqref{eq:Bessel} and \eqref{eq:parabolic}, the relevant scale that controls particle production is $H=m$; after the Hubble parameter decreases below $m$, particle production ceases and the Bogoliubov coefficient becomes constant. The scale factor at this moment is given by
\be
H(a_*) = m ~\Rightarrow ~ a_* = a_r\left(\frac{H_r}{m}\right)^{\frac{2}{3(1+\omega)}},
\ee
which can be satisfied either during reheating or radiation domination.

Consider first the case in which $H=m$ happens during reheating. We need simply to take the solutions in Eq.\,\eqref{eq:solution_RH}, compute the Bogoliubov coefficient according to Eq.\,\eqref{eq:Bogoliubov} and integrate over $k$, resulting in
\al{\label{eq:na3_analytical_RH}
na^3 = \int_0^\infty\frac{\dd k}{2\pi^2}\, k^2|\beta_k|^2\Big|_{a=a_*} = \frac{(a_eH_e)^3}{1440\pi^2}\frac{m}{H_e}\Bigg[19+20\left(\frac{m}{H_e}\right)^3&-9\left(\frac{m}{H_e}\right)^{10/3}\Bigg],\\
&H=m\text{~during~reheating}.
}
As expected, we see that $na^3$ is proportional to the mass, the only parameter that breaks conformal symmetry in our case.

For $H=m$ during radiation domination, we instead use the solutions computed in Eq.\,\eqref{eq:solution_rad}, and the corresponding number density is
\al{\label{eq:na3_analytical_Rad}
na^3 & = \frac{(a_{\rm RH}H_{\rm RH})^3}{1440\pi^2}\frac{m}{H_{\rm RH}}\Bigg\{18\left(\frac{m}{H_{\rm RH}}\right)^{1/2}+2\left(\frac{m}{H_{\rm RH}}\right)^{3}-\left(\frac{m}{H_{\rm RH}}\right)^{7/2}-\\
&\qquad -9\left(\frac{m}{H_{e}}\right)^{10/3}\left(\frac{m}{H_{\rm RH}}\right)^{1/6}+10\left(\frac{m}{H_{e}}\right)^{3}\left[1+\left(\frac{m}{H_{\rm RH}}\right)^{1/2}\right]\Bigg\},\\
& \qquad\qquad\qquad \qquad\qquad\qquad \qquad\qquad\qquad H=m~\text{during radiation domination}.
}
Similar to Eq.\,\eqref{eq:na3_analytical_RH}, the number of particles is proportional to $m$. Also notice that the expression is sensitive to both $H_e$ and $H_\text{RH}$.

In Fig.\,\ref{fig:scalar_analytical} we compare the results obtained in Eqs.\,\eqref{eq:na3_analytical_RH} and \eqref{eq:na3_analytical_Rad} with full numerical results (see Sec.\,\ref{sec:results}) for $H_e=10^{13}$ GeV and $T_\text{RH}=10^{12}$ GeV. We see that up to $m\lesssim H_e$ the agreement is incredibly good, proving that our strategy of separating the regions $k\ll a(\eta) m$ and $k\gg a(\eta)m$ is reasonable. For masses larger than the Hubble scale at the end of inflation, the assumptions we used to obtain the solutions break down and the analytical approximation becomes unreliable. From the figure, this point becomes clear, as the numerical solution falls off exponentially while the analytical approximation diverges.
%%%%%%%%%%%%%%%%%%%%%%%%%%%%%%%%%%%%%%%%%%%%%%%%
\begin{figure}[t!]
\centering
    \includegraphics[width=0.9\linewidth]{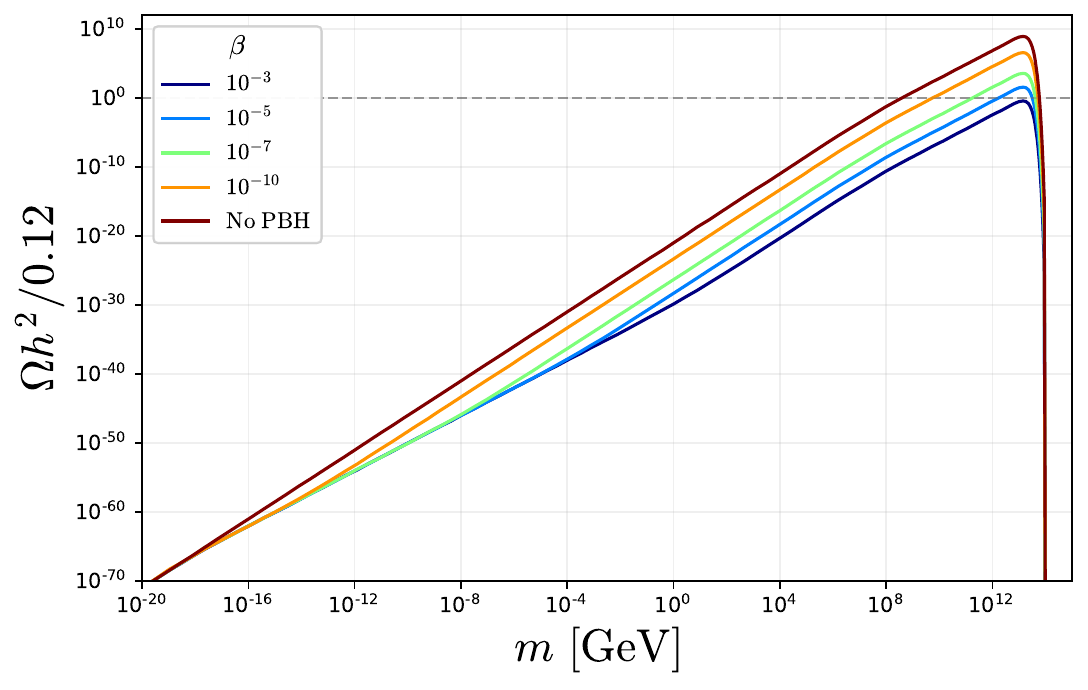}
    \caption{Abundance of a conformally coupled scalar ($\xi = 1/6$) as a function of its mass $m$ for various values of $\beta$. We consider only the DM population produced by CGPP, taking into account the effects of entropy dilution and the additional phase of matter domination due to the presence of a PBH population. We do not add to the abundance the DM population produced during the BH evaporation.}
    \label{fig:conformal_scalar_Oh2}
\end{figure}
%%%%%%%%%%%%%%%%%%%%%%%%%%%%%%%%%%%%%%%%%%%%%%%%

\subsection{Numerical results with PBHs}

We can now solve Eq.\,\eqref{eq:EoM_scalar_app} numerically including also an initial population of PBHs. Fixing $H_e=10^{13}$ GeV and $T_\text{RH}=10^{12}$ GeV, that corresponds to $M_\text{BH}=1.8\times 10^7~\text{g}$, we show in Fig.\,\ref{fig:conformal_scalar_Oh2} the total abundance of gravitationally produced DM for different values of $\beta$. In the plot we take into account the additional phase of matter domination and the entropy dilution due to the PBH population, but we do not add the DM population produced by BH evaporation. As already commented in Sec.\,\ref{sec:results}, in this case the difference between masses that reach $H=m$ during matter or radiation domination is much less pronounced with respect to the cases shown in Fig.\,\ref{fig:abundance}. On the other hand, the effect of entropy dilution is clear and diminshes the final abundance by many orders of magnitude. As already observed in Fig.\,\ref{fig:abundance}, for very small masses the presence of the PBH population is irrelevant, since for such small masses the final abundance is set well after all the dynamics related to PBHs has stopped.

%%%%%%%%%%%%%%%%%%%%%%%%%%%%%%%%%%%%%%%%%%%%%%%%%%%%%%%%%%%%%%%%%%%%%%%%%%%%%%%%%%%%%%%%%%
\section{Power spectrum for inflationary curvature perturbations}\label{app:power_spectrum}
%%%%%%%%%%%%%%%%%%%%%%%%%%%%%%%%%%%%%%%%%%%%%%%%%%%%%%%%%%%%%%%%%%%%%%%%%%%%%%%%%%%%%%%%%%

In what follows, we provide a short summary of the derivation of the Mukhanov--Sasaki equations, the numerical procedure for deriving its solutions, and the computation of the comoving power spectrum, the basic ingredient for obtaining the initial PBH energy density fraction, see Eq.\,\eqref{eq:beta_in}.

Starting from a slow-roll model of inflation, one can consider the scalar, \textit{i.e.}, comoving curvature, gauge-invariant perturbations ${\cal R}$ arising from the excitation during inflation of metric fluctuations\,\cite{Baumann:2009ds}.
The action for slow-roll inflation
\begin{align}
    S=\frac{1}{2}\int d^4x \, \sqrt{-g}\,[g^{\mu\nu} \partial_\mu \Phi \partial_\nu \Phi -  2V(\Phi) + m_{\rm PL}^2 R]\, ,
\end{align}
is expanded to the second order in ${\cal R}$ to obtain, see Ref.\,\cite{Baumann:2009ds} for further details,
\begin{align}
    S_{(2)} = \frac{1}{2} \int d^4x\, a^3\, \frac{\dot{\Phi}^2}{H^2} (\dot{\cal R}^2 - a^{-2} (\partial_i {\cal R})^2).
\end{align}
Defining the new variable $u=z{\cal R}$, with $z = a \dot{\Phi}/H$, we can obtain the Mukhanov--Sasaki equation for the Fourier mode $u_k$\,\cite{Sasaki:1986hm,Mukhanov:1988jd}
\begin{align}\label{eq:MS-eq}
    u_k'' + \left( k^2 - \frac{z''}{z}\right)u_k = 0.
\end{align}
We can observe the similarity of this equation with the mode equations for the scalar field produced via CGPP in Eq.\,\eqref{eq:EoM_scalar}. 
Thus, we solve the Eq.\,\eqref{eq:MS-eq} considering a Bunch--Davies initial condition for scales $k \gg a H$,
\begin{align*}
    u_k \to \frac{e^{-ik\eta}}{\sqrt{2k}}.
\end{align*}
Technically, we consider Eq.\,\eqref{eq:MS-eq} as function of $N_e = \log(a)$ to simplify the numerical approach\,\cite{Ballesteros:2017fsr}. Such equation is given by
\begin{align}
    \frac{d^2 u_k}{dN_e^2} + (1-\epsilon) \frac{du_k}{dN_e} + \left[\frac{k^2}{a^2 H^2} + (1+\epsilon-\kappa)(\kappa-2)-\frac{d(\epsilon-\kappa)}{dN_e}\right]u_k=0,
\end{align}
where $\epsilon,\kappa$ are the slow-roll parameters defined in Eq.\,\eqref{eq:slow-roll-pars}.

The evolution of the mode equations begins at an initial value of the scale factor that ensures that the mode is within the horizon. More precisely, we set the scale factor $a$ at a point four $e$-folds before the mode exits the horizon. Subsequently, we employ the Mukhanov--Sasaki equations to track the mode's evolution until the end of inflation.
At such a point, we obtain the power spectrum via\,\cite{Ballesteros:2017fsr, Mishra:2019pzq}
\begin{align}
    {\cal P}_{\cal R} = \frac{k^3}{2\pi^2}\frac{|u_k|^2}{z^2}.
\end{align}
We stop the evolution at the end of inflation since the mode is constant outside the horizon\,\cite{Baumann:2009ds}.

We employ the solutions of the Mukhanov--Sasaki equations to normalize the inflaton potential, either having an addtional feature to produce PBHs or not, to make it compatible with CMB observations\,\cite{Planck:2018jri}.
This is achieved by imposing the constraint
\begin{align}
    {\cal P}_{\cal R}(k_*) = 2.1\times 10^{-9},
\end{align}
with $k_* = 0.05~{\rm Mpc}^{-1}$ the pivot scale.
We also require that the number of $e$-folds between the largest observable scales $k \sim 10^{-4}~{\rm Mpc}^{-1}$ and the end of inflation to be in the range of $\sim 50-60$ in order to assure the solutions of the horizon and flatness problems of the Universe.
This is done by imposing the number of $e$-folds between end of inflation and the CMB scales to be $\Delta N_e \sim 45 - 55$\,\cite{Ballesteros:2017fsr}.

%%%%%%%%%%%%%%%%%
%%%	REFERENCES	   %%%		
%%%%%%%%%%%%%%%%%

\newpage
\bibliographystyle{JHEP2}
{\footnotesize
\bibliography{main}}
\end{document}